\documentclass{aa}

\usepackage{graphics}
\usepackage{amsmath}
\usepackage{amssymb}
\usepackage{natbib}
\begin{document}

\thesaurus{06(08.05.3; 08.14.1; 08.16.6; 08.18.1; 02.04.1; 02.05.2)}

\title{Properties of non-rotating and rapidly rotating protoneutron stars}

\author{Klaus Strobel \and Christoph Schaab \and Manfred K. Weigel}

\institute{Sektion Physik, Ludwig-Maximilians Universit{\"a}t,
  Am Coulombwall 1, D-85748 Garching, Germany}
  
\offprints{K. Strobel, \\
               e-mail: kstrobel@laser.physik.uni-muenchen.de}

\date{Received 15 December 1998 / Accepted 28 July 1999}

\titlerunning{Properties of non-rotating and rapidly rotating protoneutron 
stars}
\authorrunning{K. Strobel et al.}

\maketitle

\begin{abstract}

Properties of non-rotating and rapidly rotating protoneutron stars 
and neutron stars are investigated. 
Protoneutron stars are hot, lepton rich neutron stars 
which are formed in Type-II supernovae. The hot dense matter is 
described by a realistic equation of state which is obtained by 
extending a recent approach of Myers and \'Swi\c atecki to the nuclear 
mass formula.  We investigate the properties of protoneutron stars and
neutron stars at different evolutionary stages in order to emphasize
the differences between very young and old neutron stars.  The
numerical calculations are performed by means of an exact description
of rapid, uniform rotation in the framework of general relativity. We
show that the minimal marginally stable protoneutron star mass is
much higher than the corresponding minimum mass of a cold neutron
star. The minimum gravitational (baryonic) mass of 
0.89 - 1.13$\,M_{\sun}$ (0.95 - 1.29$\,M_{\sun}$) of a neutron star is
therefore determined at the earliest stages of its evolution. We also
show that the use of different temperature profiles in the envelope as
well as different shapes of the neutrino sphere change the properties
of protoneutron stars and hot neutron stars by up to 20\,\%.  A
preliminary analysis indicates that even the most massive protoneutron
stars rotating with Kepler frequency are secularly stable. Under the
assumption of conserved angular momentum and baryonic mass, the
maximum rotational frequency of an evolved neutron star is determined
by the Kepler frequency of the protoneutron star. We can thus derive
a lower limit, $P_{\rm min}\sim 1.56\,-\,2.22$~ms, to the rotational 
period of young neutron stars with a canonical gravitational mass of
$1.35\,M_{\sun}$. This result furtherly supports the assumption that
millisecond pulsars are accelerated due to accretion onto a cold
neutron star.

\keywords{Stars: evolution -- Stars: neutron -- Stars: pulsars 
          -- Stars: rotation -- Dense matter -- Equation of state}
\end{abstract}

\section{Introduction} \label{sec1}

It is generally believed that a neutron star (NS) is born as a result
of the gravitational collapse of the iron core of a massive evolved
progenitor star ($M\sim 8\,-\,25\,M_{\sun}$) in a Type-II supernova
\cite[e.g.][]{Bet90}. The iron core of such a star collapses when its
mass reaches the Chandrasekhar limit
\begin{equation}
\label{1.1}
M_\mathrm{Ch} \sim 5.8Y^{2}_\mathrm{e} M_{\sun}
              \sim 1.2 \ldots 1.5 M_{\sun}\, ,
\end{equation}
where $Y_\mathrm{e}$ denotes the number of electrons per baryon which
depends on the mass of the progenitor star. Due to the Fermi pressure of 
the nucleons, the collapse stops when nuclear matter density is reached 
and the core bounces back.  Shortly after core bounce (some
10 ms) a hot, lepton rich NS, called {\it protoneutron} star (PNS), is
formed. This PNS consists of a shocked envelope with an entropy per
baryon\footnote{$k_\mathrm{B} =
\hbar = c = 1$ throughout this paper.}  $s$ $\sim$ 4 - 10 and an
unshocked core with $s \sim 1$ \cite[]{BHF95}.  
The envelope and the core contain
nearly the same mass of about $0.6$ - $0.8 M_{\sun}$ \cite[slightly
depending on the mass of the progenitor star, see][]{BHF95, KJM96}.
During the so-called Kelvin-Helmholtz cooling phase
\cite[e.g.][]{Jan93} the lepton number decreases in the PNS due to the
loss of neutrinos and consequently the PNS evolves in about 10 to 30
seconds into a hot, lepton poor neutron star (HNS) with an entropy per 
baryon, $s\sim
1\,-\,2$, depending on the model \cite[e.g.][]{BL86, KJ95, SST95,
PRPLM98}.  After several minutes this HNS cools to a cold neutron star 
(CNS) with
temperatures $T < 1$ MeV throughout the star
\cite[e.g.][]{KJ95}.  Finally, it slowly cools via neutrino and photon
emission until its thermal radiation is too weak to be observable after
about $10^7$~yr \cite[e.g.][]{Tsuruta66,Schaab96}.

The PNS is in $\beta$-equilibrium during its lifetime, since the time
scale of the weak-interaction is much smaller than the
evolutionary time scale, i.e. the neutrino diffusion time scale or the
neutrino cooling time scale, respectively.  Hence the evolution of a
PNS can be studied by considering quasi-stationary models at different
times.

The properties of PNS's were investigated by different authors. For
example, the case of non-rotating PNS's was studied by \cite{TNH94},
\cite{Bom95}, and \cite{Pra97}.  The case of rotating PNS's was treated
by \cite{RDIMP92} and \cite{Tak95, Tak96} by means of an empirical
formula for the Kepler frequency which was developed for CNS's
\cite[e.g.][]{HSB95}. \cite{HOE94} and \cite{GHZ97} account for rapid
rotation by using an exact, general relativistic approach.  Finally,
\cite{GHZ98} have performed the case of differential rotation of PNS's.
Most of these authors did not utilize an equation of state (EOS) of
hot matter throughout the whole star \cite[except][]{RDIMP92, GHZ97,
GHZ98}, but used an EOS of cold matter for the envelope of the star
instead.  As we will show, this simplification leads to radii (Kepler
frequencies) which are too small (large).

The aim of this work is to study the properties of non-rotating and
rapidly, uniformly rotating PNS's.  We use an exact, general
relativistic approach to rapid rotation \cite[]{Schaab98}. The hot
dense matter is described by a recently devolped EOS \cite[]{Str98},
which is based on a modern parametrisation of the Thomas-Fermi
approach for finite nuclei and cold nuclear matter performed by
\cite{MS90, MS91, MS96}. We generalize this approach to hot dense matter
by taking the thermal effects on both the kinetic and the interaction 
energy into account. In this way, we
construct a set of EOS's with different profiles of the entropy per
baryon and different lepton numbers. We can follow the evolution of 
the PNS into a CNS, by means of this set, at different
evolutionary stages.  We also investigate the influence of the
location and the shape of the neutrino sphere as well as the influence
of the value of the temperature in the star's envelope. 

The paper is organized as follows.  Firstly, in Sect.~\ref{sec2} we
will briefly review the physics in the interior of PNS's and describe
the different evolutionary stages of PNS's and NS's.  Furthermore, we
discuss the location and the shape of the neutrino sphere.  
The EOS's of PNS and NS matter are described in Sect.~\ref{sec3}, where we
emphasize the influence of finite temperature and trapped lepton
number. The properties of rotating and non-rotating PNS's and NS's are
presented in Sect.~\ref{sec4}.  Finally, discussion of our results
and conclusions are given in Sect.~\ref{sec5}.

\section{Inside a protoneutron star} \label{sec2}

A PNS differs in several respects from a CNS: At the beginning of its
lifetime the PNS contains a high lepton number, $Y_\mathrm{l}$, since
the core is opaque with respect to neutrinos.  A further difference is
the high temperature which cannot be neglected with respect to the
Fermi temperature throughout the whole PNS.  We define, as usual, the
lepton number, $Y_\mathrm{l}=Y_\mathrm{e}+Y_{\nu_\mathrm{e}}$, as the
sum of the net electron fraction,
$Y_\mathrm{e}=(n_\mathrm{e^{-}}-n_\mathrm{e^{+}})/n$, (where $n$,
$n_\mathrm{e^{-}}$, and $n_\mathrm{e^{+}}$ are the baryon number
density, the electron number density, and the positron number density,
respectively) and the net electron neutrino fraction,
$Y_{\nu_\mathrm{e}}=(n_{\nu_\mathrm{e}}-n_{\bar \nu_\mathrm{e}})/n$,
(where $n_{\nu_\mathrm{e}}$ and $n_{\bar \nu_\mathrm{e}}$ are the
electron neutrino number density and the electron anti-neutrino number
density, respectively).  Since the muon number density is small in a
PNS, they are neglected here.  The reason for the
small muon number is that the muon lepton family number is conserved,
$Y_{\mu} + Y_{\nu_\mu}=0$, while the neutrinos are
trapped\footnote{This statement is also true for massive neutrinos
recently detected by the Super-Kamiokande collaboration
\cite[see][]{Fuk98}.  The detected mass is too small to allow
considerable flavour oscillations during the first seconds of a
supernova \cite[see discussion in Section 9.5.3 of][]{Raf96}}.
Relativistic calculations lead to the conclusion that NS's and PNS's
are composed not only of nucleons and leptons but also of hyperons
and, possibly, of nucleon isobars \cite[see, e.g.][]{Pan71, 
SM96, BG97, HWWS98}. 
Nevertheless, we shall not take these particle species into
account. In view of the rather large uncertainties of the hyperon
couplings, we shall, as a first approach to this problem, neglect
these additional degrees of freedom. In the following we select four
different stages in the evolution of PNS's and NS's, namely at times
$t_{1} \sim 50 - 100$ ms, $t_{2} \sim 0.5 - 1$ s, $t_{3} \sim 10-30$
s, and $t_{4} = $ some minutes after core bounce.

\subsection{Protoneutron stars about 50 - 100 ms after core bounce} 
\label{sec21}

\begin{table*}
  \caption[]{Entropies, temperatures, densities, and lepton numbers 
             used in this paper.
             The entries are: entropy per baryon or temperature in the 
             envelope, $s_\mathrm{env}$, $T_\mathrm{env}$; 
             entropy per baryon or temperature in the 
             core, $s_\mathrm{core}$, $T_\mathrm{core}$; 
             maximum baryon number density of the envelope correlated 
             with the 
             entropy per baryon or temperature in the envelope, 
             $n(s_\mathrm{env}$, $T_\mathrm{env}$); 
             minimum baryon number density of the core correlated 
             with the 
             entropy per baryon or temperature in the core, 
             $n(s_\mathrm{core}$, $T_\mathrm{core}$); 
             baryon number density below which the neutrinos are not trapped,
             $n(Y_\mathrm{l,env})$;
             baryon number density above which the neutrinos are totally trapped,
             $n(Y_\mathrm{l,core})$; 
             lepton fraction inside the core, $Y_\mathrm{l,core}$.}
  \label{EOSs}
  \begin{tabular}{ l c c c c c c c }
  \hline
   & & & & & & & \\
  Label  & $s_\mathrm{env}$, $T_\mathrm{env}$ & 
           $s_\mathrm{core}$, $T_\mathrm{core}$ & 
           $n(s_\mathrm{env}, T_\mathrm{env})$ & 
           $n(s_\mathrm{core}, T_\mathrm{core})$ & 
           $n(Y_\mathrm{l,env})$ & $n(Y_\mathrm{l,core})$ & 
           $Y_\mathrm{l,core}$ \\
       &  &  & [fm$^{-3}$] & 
                      [fm$^{-3}$] & [fm$^{-3}$] & [fm$^{-3}$] &   \\
   & & & & & & & \\
  \hline
   & & & & & & & \\
  EPNS$^\mathrm{s5s1}_\mathrm{YL04}$      & 5.0 & 1.0 & 0.02 & 0.1 & 
                                          0.0006 & 0.0006 & 
                                          0.4 \\
  EPNS$^\mathrm{s4s1}_\mathrm{YL04}$      & 4.0 & 1.0 & 0.02 & 0.1 & 
                                          0.0006 & 0.0006 & 
                                          0.4 \\
  & & & & & & & \\
  LPNS$^\mathrm{s2}_\mathrm{YL04}$        & 2.0 & 2.0 & - & - &0.0006 &
                                          0.0006 & 0.4 \\
  LPNS$^\mathrm{T06s2}_\mathrm{YL04}$     & 0.6 MeV & 2.0 & 0.0004 & 
                                          0.0004 & 
                                          0.0006 & 0.0006 & 
                                          0.4 \\
  LPNS$^\mathrm{s2}_\mathrm{YL04(64-63)}$$^*$ & 2.0 & 2.0 & - & - & 
                                          0.0006 & 0.006 & 
                                          0.4 \\
  LPNS$^\mathrm{s2}_\mathrm{YL04(64-22)}$& 2.0 & 2.0 & - & - &0.0006 & 
                                          0.02 & 0.4 \\
  LPNS$^\mathrm{s2}_\mathrm{YL04(64-62)}$& 2.0 & 2.0 & - & - &0.0006 &
                                          0.06 & 0.4 \\
  LPNS$^\mathrm{s2}_\mathrm{YL03}$        & 2.0 & 2.0 & - & - &0.0006 &
                                          0.0006 & 0.3 \\
  LPNS$^\mathrm{s1}_\mathrm{YL04}$        & 1.0 & 1.0 & - & - & 0.0006 & 
                                          0.0006 & 0.4 \\
  LPNS$^\mathrm{s1}_\mathrm{YL03}$        & 1.0 & 1.0 & - & - & 0.0006 & 
                                          0.0006 & 0.3 \\
  LPNS$^\mathrm{s0}_\mathrm{YL04}$        & 0 & 0  & - & - & 
                                          0.0006 & 0.0006  & 0.4  \\
  & & & & & & & \\
  HNS$^\mathrm{s2}$                       & 2.0 & 2.0 & - & - & - & - & - \\
  HNS$^\mathrm{T06s2}$                    & 0.6 MeV & 2.0 & 0.0004 & 
                                          0.0004 & - & - & - \\
  HNS$^\mathrm{T0s2}$                     & 0 MeV & 2.0 & $0.06$ & 
                                          $0.06$ & - & - & - \\
  HNS$^\mathrm{s1}$                       & 1.0 & 1.0 & - & - & - & - & -  \\
  HNS$^\mathrm{T03s1}$                    & 0.3 MeV & 1.0 & 0.0004 & 
                                          0.0004 & - & - & - \\
  HNS$^\mathrm{T0s1}$                     & 0 MeV & 1.0 & $0.06$ & 
                                          $0.06$ & - & - & - \\
  & & & & & & & \\
  CNS                                     & 0 MeV & 0 MeV & - & - & - & - 
                                                                  & -  \\
   & & & & & & & \\
  \hline
  \end{tabular} \\
  * This notation means that the first number in parentheses 
  classifies the lower density
  boundary of the neutrino sphere, i.e. $6 \times 10^{-4}$, the second
  number gives the upper density boundary, i.e. $6 \times 10^{-3}$.
\end{table*}
This early type protoneutron star (EPNS) is characterized by a hot
shocked envelope with an entropy per baryon of $s \sim $ 4 - 5 for
densities $n<0.02$ fm$^{-3}$, an unshocked core with $s \sim 1$ for
densities $n>0.1$ fm$^{-3}$, and a transition region between these
densities \cite[]{BHF95}.  The entropy per baryon in the very outer
layers of an EPNS is larger than $s \sim 10$. However, these layers
have only a small influence on the EPNS structure and are therefore
neglected.  We investigate EPNS models with constant lepton number,
$Y_\mathrm{l} = 0.4$ for densities above $n = 6 \times 10^{-4}$
fm$^{-3}$ where the neutrinos are trapped \cite[]{BHF95}. Below this
density, the neutrinos can freely escape and the chemical potential of
the neutrinos vanishes, $\mu_{\nu_\mathrm{e}} = 0$
\cite[]{Coo88}. We refer to Table~\ref{EOSs} for the detailed
parameters of the EPNS models studied here.

\subsection{Protoneutron stars at $t \sim 0.5 - 1$ s after core bounce}
\label{sec22}

At this later stage, the entropy per baryon is approximately constant
throughout the star, $s \sim 2$, except in some outer regions
\cite[]{BL86, KJ95, Kei96}.  The lepton number is approximately constant 
since the neutrino diffusion time, $\tau_{\nu-\rm{diff}}\sim 10$~s, is
by an order of magnitude larger than the PNS's age. We model this
late type protoneutron star (LPNS) with a neutrino transparent
envelope with densities $n < n_\mathrm{env} = 6 \times 10^{-4}$
fm$^{-3}$ and a neutrino opaque core with densities $n >
n_\mathrm{core}$ and $Y_\mathrm{l} = 0.4$ (see Table~\ref{EOSs}).  The
transition region between $n_\mathrm{env}$ and $n_\mathrm{core}$ is
called \emph{neutrino sphere} \cite[]{Jan93}.  We choose four
different values for $n_\mathrm{core}$ to simulate the influence of
the shape of the neutrino sphere on the structure of LPNS's.  We show
also a LPNS with $Y_\mathrm{l} = 0.3$ for densities $n > 6 \times
10^{-4}$ fm$^{-3}$, for the sake of comparison, since different
evolution calculations show different lepton numbers \cite[]{BL86,
KJ95, PRPLM98}.

These different choices are motivated by recent theoretical
calculations of the neutrino-nucleon cross-section that include
modifications due to the nucleon-nucleon interaction and spin-spin
correlations. The problem was treated, for instance, by \cite{Raf96},
\cite{RPL98}, \cite{BS98}, and \cite{PRPLM98} using static correlation 
functions and by \cite{Raf96}, \cite{JKRS96}, \cite{RS97}, and 
T. Strobel (work in preparation)
by using dynamical correlation functions. Because of the
high complexity of the problem, the behaviour of the cross section and
thus of the location and shape of the neutrino sphere is rather
uncertain.
Another reason for the different choices is the uncertainty due to 
convection, that might considerably influence the cooling of PNS's 
\cite[e.g.][]{BL88, KJM96, Mez98}.

For the sake of comparison, we investigate also a model with an
isothermal envelope with
temperature\footnote{We choose $T$ and not $T^*$ for simplicity,
because the values of $T$ and $T^*$ are comparable in the envelope;
for the definition of $T^*$, which includes metric corrections, see
\cite{GHZ97} and \cite{Gon97}.} $T = 0.6$ MeV
for densities below $n \sim 4 \times 10^{-4}$ fm$^{-3}$. This
temperature value is motivated by the fact that the temperature in the
central parts of the progenitor star approximately raises to this value
before the onset of the core collapse \cite[e.g.][]{ST83, Bet90}. It
is certainly an upper limit of the true temperature since it
corresponds to an increase of the entropy per baryon by three or four
orders of magnitude for densities $n \sim 6 \times 10^{-10}$
fm$^{-3}$ (outer most layer of the star).  This high
entropy seems to be possible only in hot bubbles \cite[see][]{MWW93}.

For comparison we investigate also an unphysical, cold EOS, 
LPNS$^\mathrm{s0}_\mathrm{YL04}$, with a trapped 
lepton number of $Y_\mathrm{l} = 0.4$ for densities 
$n > 6 \times 10^{-4}$ fm$^{-3}$.

\subsection{Deleptonized hot neutron star at $t \sim 10 - 30$ s 
            after core bounce} \label{sec23}

After $10 - 30$~s the neutrinos can freely escape and the HNS is
nearly deleptonized. This also means that the lepton family number is
not conserved anymore. The $\beta$-equilibrium is thus given by
$\mu_\mathrm{p} + \mu_\mathrm{e} = \mu_\mathrm{n}$, $\mu_\mathrm{e} =
\mu_{\mu}$, and $\mu_{\nu} = 0$ for all neutrino species.  At this
stage muons have to be taken into account since the muon number
density is comparable to the electron number density above nuclear matter
density.  The entropy per baryon is nearly constant, $s\sim 1-2$,
during the evolution from the LPNS to the HNS
\cite[e.g.][]{BL86, KJ95, SST95, PRPLM98}.

We again compare the models with isentropic envelopes with models with
isothermal envelopes, $T = 0.3$ MeV or $T = 0.6$ MeV for
densities below $n \sim 4 \times 10^{-4}$ fm$^{-3}$. The models for
the deleptonized HNS's are summarized in Table \ref{EOSs}.

\subsection{Cold neutron star some minutes after core bounce}
\label{sec24}

After some minutes the NS has a temperature of $T < 1$ MeV throughout
the star and the EOS for cold NS matter can be used to describe the
CNS, because the thermal effects are negligibly small
\cite[see][]{ST83}.  We shall adopt the model derived by \cite{BPS71} 
for densities below neutron drip density, $n<2.6\times
10^{-4}$~fm$^{-3}$, and \cite{NV73} for densities between neutron
drip density and the transition density, $2.6\times
10^{-4}<n<0.1$~fm$^{-3}$. Above this transition density, we use the
model for CNS matter in $\beta$-equilibrium without neutrinos derived
by \cite{Str97}.

\section{Equation of state for protoneutron stars}
\label{sec3}

The EOS of PNS matter is the basic input quantity whose knowledge over
a wide range of densities, ranging from the density of iron at the
star's surface up to about eight times the density of normal nuclear
matter reached in the cores of the most massive stars of a sequence, is
necessary to solve the structure equations.  Due to the high lepton
number, the EOS of PNS's is different from the EOS's for cold and hot
NS's with low lepton numbers.  The nuclear EOS used in this paper for
the description of a PNS is a Thomas-Fermi model of average nuclear
properties, with a momentum- and density-dependent, effective
nucleon-nucleon interaction developed by \cite{MS90, MS91}.  The
parameters of the nuclear EOS were adjusted to reproduce a wide range
of properties of normal nuclear matter and nuclei \cite[]{MS95, MS96,
MS98, WCS97}. \cite{Str98} extended this 
approach to the case of finite
temperature\footnote{The EOS's used in this paper are available at: 
http:\hspace{0cm}//www.\hspace{0cm}physik.\hspace{0cm}uni-muenchen.\hspace{0cm}de\hspace{0cm}/sektion\hspace{0cm}/suessmann\hspace{0cm}/astro.}, 
where they use exact numerical solutions for the integration 
over the Fermi-Dirac distribution functions.
Appendix A contains a brief description of this approach.  
We extend the nuclear EOS to subnuclear
densities and to different compositions of PNS and HNS matter
(i.e. trapped neutrinos, constant entropy per baryon, $\ldots$).  In
the subnuclear regime, the EOS was also obtained by means of the
homogeneous Thomas-Fermi model. The electron number is derived by
fitting the pressure to the subnuclear EOS's of \cite{BPS71} and
\cite{NV73} for densities below the density of the neutrino sphere in
the case of EPNS and LPNS models and below the nuclear density in the
case of HNS models.  Our results for subnuclear densities are
comparable to the EOS's derived by
\cite{LS91} and used in the investigations of
\cite{GHZ97, GHZ98} and \cite{Gon97}. 

\begin{figure}
  \resizebox{\hsize}{!}{\rotatebox{-90}{\includegraphics{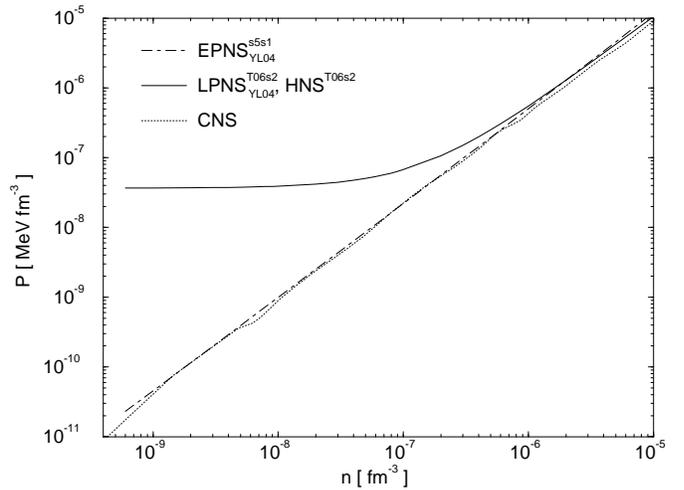}}}
  \caption[]{Pressure versus baryon number density for densities 
             $n < 10^{-5}$ fm$^{-3}$ of hot dense matter.
             The curve CNS corresponds to cold matter (see 
             Sect. \ref{sec24}). The 
             LPNS$^\mathrm{T06s2}_\mathrm{YL04}$ and 
             HNS$^\mathrm{T06s2}$ curve is the isothermal
             envelope part of an isentropic core with entropy
             per baryon $s = 2$ (see Sects. \ref{sec22} and  
             \ref{sec23} for explanation). Finally the curve
             EPNS$^\mathrm{s5s1}_\mathrm{YL04}$ corresponds to
             the envelope part of this EOS with $s=5$ and
             no trapped lepton number in this density region.}
  \label{pn1a}
\end{figure}
\begin{figure}
  \resizebox{\hsize}{!}{\rotatebox{-90}{\includegraphics{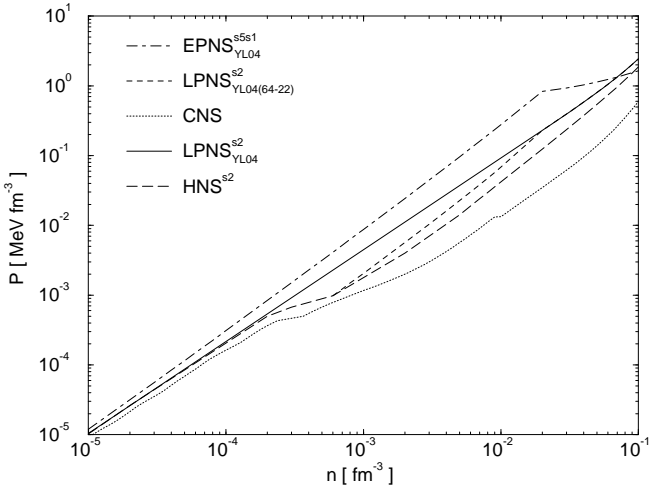}}}
  \caption[]{Pressure versus baryon number density in the density region  
             $10^{-5}$ fm$^{-3}< n < 10^{-1}$ fm$^{-3}$ for different EOS's
             of hot dense matter. The abbreviations are 
             described in Table~\ref{EOSs}. }
  \label{pn1b}
\end{figure}
\begin{figure}
  \resizebox{\hsize}{!}{\rotatebox{-90}{\includegraphics{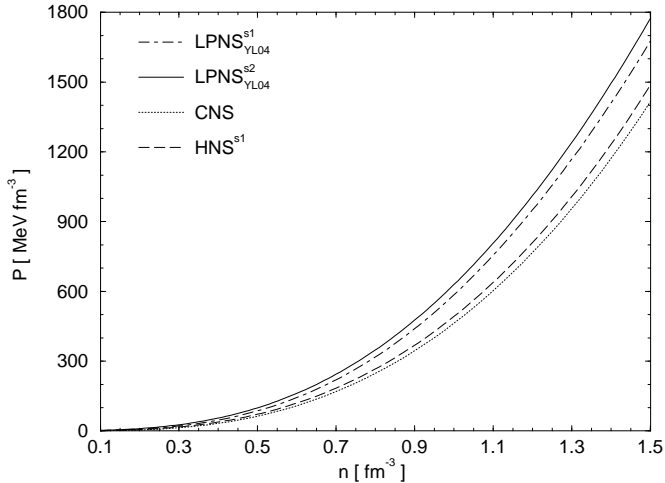}}}
  \caption[]{Pressure versus baryon number  density for densities
             $n > 0.1$ fm$^{-3}$  for different EOS's of our model 
             of hot dense matter. The abbreviations are
             described in Table~\ref{EOSs}.
             The pressure of the HNS$^\mathrm{s2}$ EOS is nearly 
             identical to the LPNS$^\mathrm{s1}_\mathrm{YL04}$ case in 
             this density region and is not shown for that reason.}
  \label{pn1c}
\end{figure}
Figures \ref{pn1a} - \ref{pn1c} show the pressure, $P(n)$, as function
of the baryon number density, $n$, for different physical scenarios.
The envelope region with densities, $n \leq 10^{-5}$~fm$^{-3}$ is
depicted in Fig. \ref{pn1a} for the isothermal part with $T =
0.6$ of the LPNS$^\mathrm{T06s2}_\mathrm{YL04}$ EOS and the
HNS$^\mathrm{T06s2}$ EOS, the isentropic
EPNS$^\mathrm{s5s1}_\mathrm{YL04}$ EOS, and the EOS for a CNS.  The
neutrinos do not contribute to the pressure for densities $n < 6
\times 10^{-4}$~fm$^{-3}$, since they are not trapped in this
region. In the EOS with $T = 0.6$~MeV, the pressure is dominated
by the contribution of the photons, $p_\mathrm{Ph} = 2.817
\times 10^{-7}\,T^4$\,MeV$^{-3}$\,fm$^{-3}$ = $3.65
\times 10^{-8}$\,MeV\,fm$^{-3}$, for low densities. 
Nevertheless, the
influence of this low density region on gross properties of LPNS's and
HNS's is only small (see Sect.\ \ref{sec4}). In contrast to the
isothermal EOS, the isentropic EOS is almost identical with the cold
EOS in this density region.

Figure \ref{pn1b} shows the general trend that the leptons dominate
the pressure in the density region around and above the neutrino
sphere, $10^{-4}$~fm$^{-3} < n < 10^{-1}$~fm$^{-3}$ for LPNS's.  Both
thermal effects and the trapped lepton number contribute significantly
to the pressure and increase it by a factor of $\sim 3 - 4$ each at
$n=0.01$~fm$^{-3}$. The thermal effects are even larger for the high
entropy model labeled EPNS$^\mathrm{s5s1}_\mathrm{YL04}$. The impact
of the shape of the neutrino sphere can be inferred by comparison of
the curves labeled LPNS$^\mathrm{s2}_\mathrm{YL04}$ and
LPNS$^\mathrm{s2}_\mathrm{YL04(64-22)}$.

With increasing density the temperature dependence of the nuclear EOS
increases. At the highest densities possible in PNS's ($n \sim 1 -
1.2$~fm$^{-3}$) the pressure increase due to thermal effects and due
to high lepton numbers become comparable (see Fig.~\ref{pn1c}).  This
feature is clearly expressed by the nearly identical pressure of the
HNS$^\mathrm{s2}$ and the LPNS$^\mathrm{s1}_\mathrm{YL04}$ EOS's. 

The EOS of \cite{LS91} shows a smaller temperature dependence at high 
densities \cite[see][]{GHZ97} in comparison with our EOS. The reason for 
this is that the temperature dependence of the EOS of \cite{LS91} lies 
entirely in the kinetic part of the enegy density, since they choose
$m^* = m$ in their approach\footnote{See Sect. 2.3 of \cite{LS91},
espacially Eqs. 2.8, 2.13, and 2.18. This will change if $m^*$ is chosen to be 
smaller than $m$.}. At this point it should be mentioned, that the behaviour 
of $m^*$ at high densities is highly uncertain. At $n\sim0.5$~fm$^{-3}$ 
we obtain, for instance, a pressure increase due to thermal effects of 
$28\,\%$ (from the LPNS$^\mathrm{s0}_\mathrm{YL04}$ to the 
LPNS$^\mathrm{s2}_\mathrm{YL04}$ case), whereas \cite{LS91} obtain an
increase of only $8\,\%$ \cite[see Fig. 1.b in][]{GHZ97}.
The impact on the structure of PNS's and
HNS's are discussed in Section \ref{sec4}, where we compare our
results with the results of \cite{GHZ97, GHZ98} and \cite{Gon97} 
who used the EOS of \cite{LS91}.

The number density and the mean energy of the neutrinos $y = \nu_{e}$,
$\bar \nu_{e}$, $\nu_{\mu}$, $\bar \nu_{\mu}$, $\nu_{\tau}$, $\bar
\nu_{\tau}$ are given by\footnote{The recently detected mass
of neutrinos \cite[see][]{Fuk98} is much smaller than the mean energy 
of the neutrinos. A finite neutrino mass therefore can be neglected.}:
\begin{equation}
\label{nany}
n_{y} = \frac{g}{2 \pi^2} \int_{0}^{\infty} 
 \frac{p^2}{ 1 + \exp (\frac{1}{T}(p - \mu_{y})) } \mathrm{d}p 
\end{equation}
and
\begin{equation}
\label{uany}
u_{y} = \frac{g}{2 n_{y} \pi^2} 
                          \int_{0}^{\infty} 
 \frac{p^3}{ 1 + \exp (\frac{1}{T}(p - \mu_{y})) } \mathrm{d}p,
\end{equation}
respectively. The trapped electron neutrinos and anti-neutrinos are in
chemical equilibrium, $\mu_{\bar
\nu_\mathrm{e}}=-\mu_{\nu_\mathrm{e}}$. The chemical potentials of all
other neutrino species vanish, $\mu_{x} = 0$ (with $x = \nu_{\mu}$,
$\bar \nu_{\mu}$, $\nu_{\tau}$, $\bar \nu_{\tau}$) due to the lepton
family number conservation (the muon number density is small compared to 
the electron number density in PNS's and 
is therefore neglected, for simplicity). The factor $g$ denotes the spin-degeneracy
factor and is related to the spin, $\vec s$, of the particles by $g =
2|\vec s| + 1$. Since only positive helicity neutrinos and negative
helicity anti-neutrinos exist\footnote{No negative helicity neutrinos
nor positive helicity anti-neutrinos were found in experiment until
now; CP violation of the weak interaction.}, the degeneracy factor is
equal 1 for the neutrinos.  In the case of vanishing chemical
potential Eqs. (\ref{nany}) and (\ref{uany}) lead to a temperature
dependence of the number density:
\begin{equation}
\label{rhonyx}
n_{x} = 1.19 \times 10^{-8}\,T^3~\mathrm{MeV}^{-3}~\mathrm{fm}^{-3} 
\end{equation}
and a linear temperature dependence of the mean neutrino energy:
\begin{equation}
\label{unyx}
u_{x} = 3.15\,T ,
\end{equation}
with $x = \nu_{\mu}$, $\bar \nu_{\mu}$, $\nu_{\tau}$, $\bar
\nu_{\tau}$.  Due to the high chemical potential of the electron
neutrinos in the case of a high trapped lepton number
($\mu_{\nu_\mathrm{e}} \gg T$; see Fig. \ref{my} and Fig. \ref{T}),
the number density and the mean energy of the electron anti-neutrinos
can be approximated by:
\begin{figure}
  \resizebox{\hsize}{!}{\rotatebox{-90}{\includegraphics{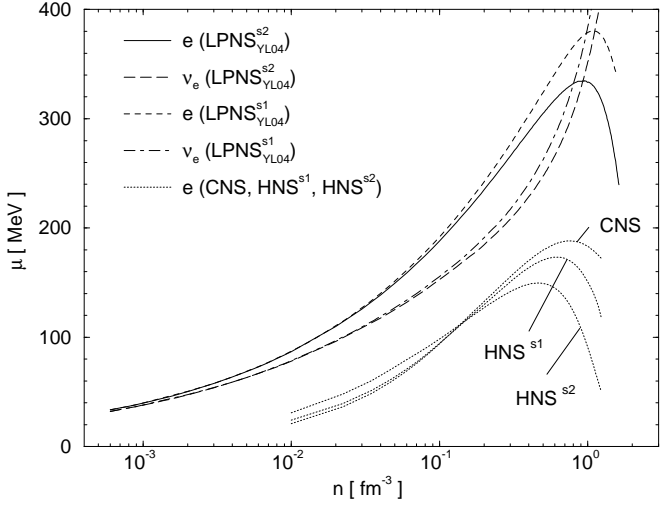}}}
  \caption[]{Chemical potential versus baryon number density
             for electrons and electron-neutrinos 
             of the LPNS$^\mathrm{s1}_\mathrm{YL04}$  and 
             LPNS$^\mathrm{s2}_\mathrm{YL04}$ cases.
             The dotted lines correspond to the CNS, 
             the HNS$^\mathrm{s1}$ and the HNS$^\mathrm{s2}$ EOS's.}
  \label{my}
\end{figure}
\begin{figure}
  \resizebox{\hsize}{!}{\rotatebox{-90}{\includegraphics{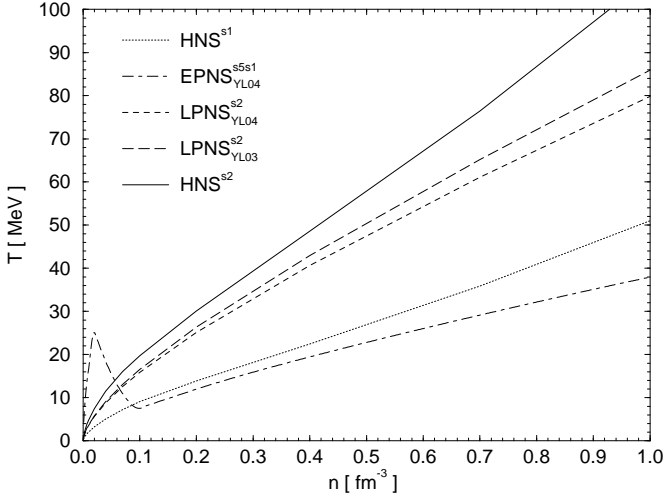}}}
  \caption[]{Temperature versus baryon number density 
             for different EOS's.}
  \label{T}
\end{figure}
\begin{equation}
\label{3.1}
n_{\bar \nu_\mathrm{e}} = \frac{T^{3}}{\pi^{2}} 
\exp({- \frac{\mu_{\nu_\mathrm{e}}}{T}}) 
\end{equation}
and
\begin{equation}
\label{3.2}
u_{\bar \nu_\mathrm{e}} = \frac{3T^{4}}{n_{\bar \nu_\mathrm{e}} \pi^{2}} 
                          \exp({- \frac{\mu_{\nu_\mathrm{e}}}{T}}) .
\end{equation}
This leads to the simple, linear temperature dependence of the mean
energy of the electron anti-neutrinos:
\begin{equation}
\label{3.3}
u_{\bar \nu_\mathrm{e}} = 3\,T ,
\end{equation}
the energy of an ultra-relativistic Boltzmann gas.
The mean neutrino energies of all neutrino species are shown in
Fig. \ref{Eny} for the case LPNS$^\mathrm{s2}_\mathrm{YL04}$. 
Figure \ref{my} shows the chemical potential of the electron neutrinos
and the electrons for different models of the EOS.
\begin{figure}
  \resizebox{\hsize}{!}{\rotatebox{-90}{\includegraphics{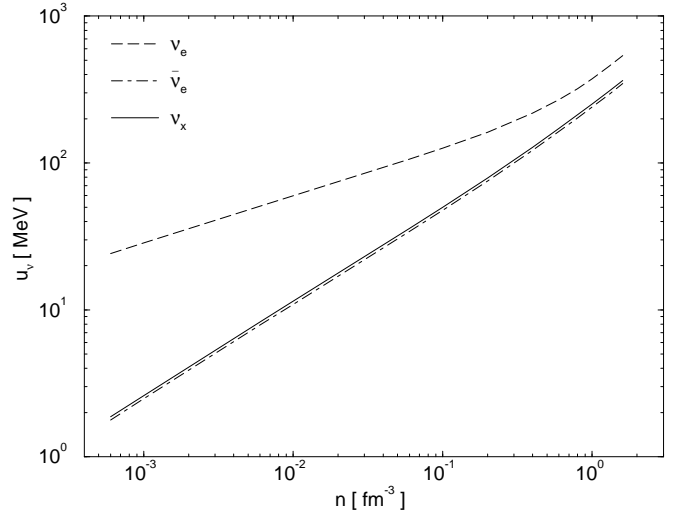}}}
  \caption[]{Mean neutrino energies versus baryon number density 
             for all neutrino types in the case
             LPNS$^\mathrm{s2}_\mathrm{YL04}$ for densities larger than
             $6 \times 10^{-4}$ fm$^{-3}$
             ($\nu_x = \nu_\mu, \bar\nu_\mu, \nu_\tau$ and $\bar\nu_\tau$).}
  \label{Eny}
\end{figure}

In Fig. \ref{T} we show the density dependence of the temperature for
different EOS's. One can see the temperature drop in the
EPNS$^\mathrm{s5s1}_\mathrm{YL04}$ EOS at the interface between the
hot shocked envelope ($n<0.02$ fm$^{-3}$) and the unshocked core
($n>0.1$ fm$^{-3}$)\cite[see][]{BL86, BHF95}. 
The temperature increases with increasing entropy per baryon 
(see discussion in Sect.~\ref{ssec:evol.seq})
and decreases with increasing lepton number, since the neutron
fraction and the proton fraction become more equal
\cite[see][]{Pra97}.  The maximum temperature in the most massive
PNS's and HNS's reaches values between 80 and 120 MeV (see
Fig.~\ref{T} and Table~\ref{max_bar_mass}).  The temperature in PNS's
and HNS's with a typical baryonic mass of $1.5 M_{\sun}$ has values
between 20 and 40 MeV (see Table~\ref{neutrinosphere}).

\begin{figure}
  \resizebox{\hsize}{!}{\rotatebox{-90}{\includegraphics{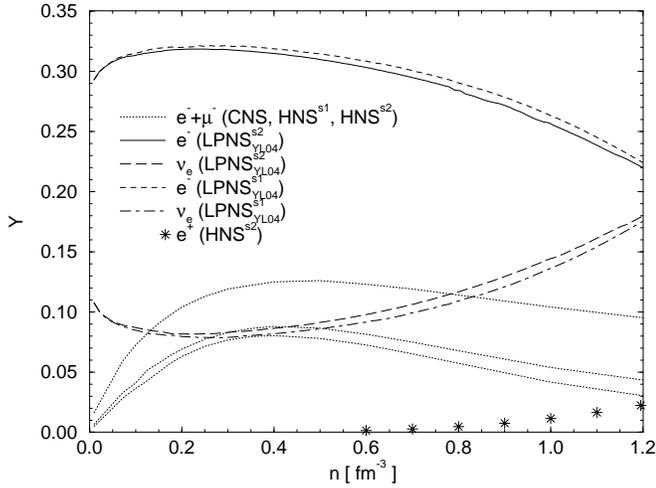}}}
  \caption[]{Lepton fractions versus baryon number density 
             for different EOS's. The lepton fractions of the 
             unphysical LPNS$^\mathrm{s0}_\mathrm{YL04}$ EOS are 
             nearly identical 
             to that of the LPNS$^\mathrm{s1}_\mathrm{YL04}$ EOS 
             and is not shown for that reason.
             The lower dotted line corresponds to 
             the CNS EOS, the middle dotted line to the 
             HNS$^\mathrm{s1}$ EOS and the upper dotted line to the 
             HNS$^\mathrm{s2}$ EOS.
             The dotted lines show the sum of the electron and muon
             fraction. The stars correspond to the positron fraction
             of the HNS$^\mathrm{s2}$ case. The positron fraction of all 
             other EOS's lies below the resolution of this figure.}
  \label{Y}
\end{figure}
The fractions of electrons and electron-neutrinos are shown in
Fig.~\ref{Y} for the LPNS$^\mathrm{s1}_\mathrm{YL04}$ and
LPNS$^\mathrm{s2}_\mathrm{YL04}$ cases with constant lepton number. 
The lepton fractions of the unphysical LPNS$^\mathrm{s0}_\mathrm{YL04}$ 
EOS are nearly identical to that of the LPNS$^\mathrm{s1}_\mathrm{YL04}$ 
EOS and are not shown in Fig.~\ref{Y} for that reason.
The fractions are nearly constant ($29 - 32\%$ for electrons and $8 -
11\%$ for electron neutrinos) in a wide range of densities.  The
electron fraction decreases for $n > 0.7$~fm$^{-3}$ since the symmetry
energy decreases for densities $n > 0.4$~fm$^{-3}$ \cite[]{Str97}.
Non-relativistic EOS's derived within variational approaches behave
similar in this respect, whereas the symmetry energy derived in
relativistic and non-relativistic Br\"uckner-Bethe calculations
monotonically increases with density \cite[see][]{Str97}.  Higher
temperatures cause a small decrease (increase) of the electron
(electron neutrino) fraction \cite[see][]{TNH94}.  Without trapped
neutrinos the sum of the electron and muon fraction increases, in
first approximation, quadratically with increasing temperature
\cite[see Fig. \ref{Y} and][]{KJ95}.

\begin{figure}
  \resizebox{\hsize}{!}{\rotatebox{-90}{\includegraphics{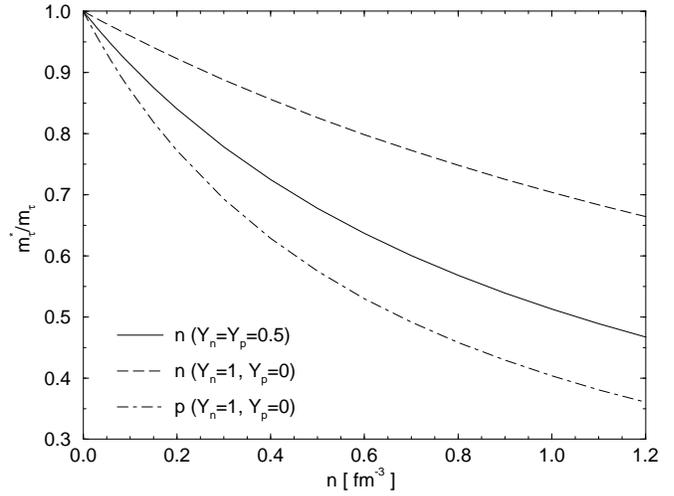}}}
  \caption[]{Effective mass of neutrons and protons versus baryon number 
             density. The solid line shows the effective mass of neutrons 
             and protons for symmetric nuclear matter (the difference 
             in the values for neutrons and protons are negligibly 
             small). The long dashed line shows the effective mass 
             of neutrons in pure neutron matter. The dot-dashed line shows 
             the effective mass of protons brought into pure neutron 
             matter.}
  \label{effmass}
\end{figure}
The effective masses, $m^*_\tau / m_\tau$, of neutrons and protons for 
symmetric and pure neutron matter are shown in Fig.~\ref{effmass}.
The effective mass for symmetric nuclear matter at saturation density is 
found to be $m^*_\tau / m_\tau = 0.867$ for neutrons and protons in our 
model. This is in good agreement with the generally accepted, experimental 
value \cite[e.g.][]{BHHQ82}. 
In the case of pure neutron matter, the effective neutron mass 
increases up to $m^*_\mathrm{n} / m_\mathrm{n} = 0.935$, at nuclear matter 
density, whereas the effective proton mass in pure neutron matter 
decreases to $m^*_\mathrm{p} / m_\mathrm{p} = 0.808$.

\begin{figure}
  \resizebox{\hsize}{!}{\rotatebox{-90}{\includegraphics{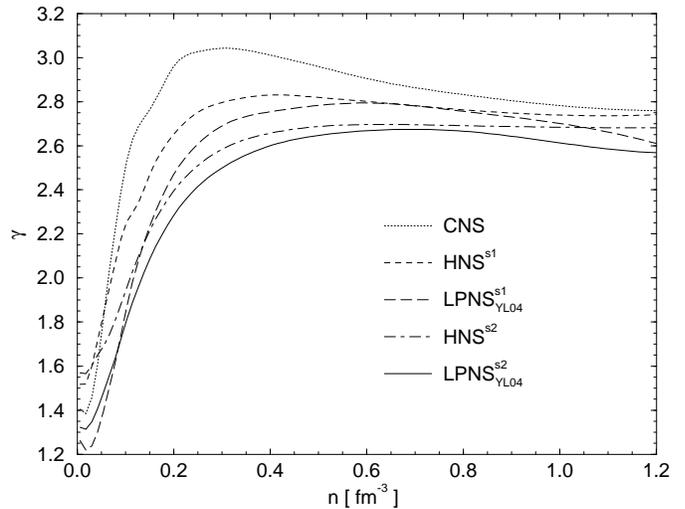}}}
  \caption[]{Adiabatic index versus baryon number density
             for different EOS's.}
  \label{gamma}
\end{figure}
The adiabatic index $\gamma$:
\begin{equation}
\label{3.4}
\gamma = \left. \frac{\mathrm{d}\ln P}{\mathrm{d}\ln n} \right|_{s}
       = \left. \frac{n}{P} \frac{\mathrm{d}P}{\mathrm{d}n} \right|_{s},
\end{equation}
is shown for different EOS's in Fig.~\ref{gamma}.  The adiabatic index
decreases with increasing temperature and lepton numbers for densities
around and above nuclear density.  In contrast, it decreases with
increasing lepton number for densities $n < 0.05$~fm$^{-3}$
\cite[see][]{Gon97}.  The steep behaviour of the adiabatic index
$\gamma$ around nuclear matter density is the reason for the core
bounce of the collapsing iron core of the progenitor star
\cite[]{ST83}.

\begin{table*}
  \caption[]{The speed of sound $v_\mathrm{s}$ in the density region
             around and above nuclear matter density for different EOS's.
             The abbreviations are described in Table~\ref{EOSs}.
             The maximum value is reached in the CNS EOS:
             $n^\mathrm{max}_\mathrm{c} (\mathrm{CNS}) = 1.246$ fm$^{-3}
             \rightarrow v_\mathrm{s}=0.964$ (in units of $c$).}
  \label{soundspeed}
  \begin{tabular}{ l c c c c c c c c c c }
  \hline
   & & & & & & & & & & \\
   & & & & & $n$ [fm$^{-3}$] & & & & & \\
   & & & & & & & & & & \\
   \cline{2-11}
    & & & & & & & & & & \\
   EOS & $0.02$ & $0.06$ & $0.1$ & $0.2$ & $0.3$ & $0.4$ & 
         $0.6$ & $0.8$ & $1.0$ & $1.2$  \\
   & & & & & & & & & & \\
  \hline
   & & & & & & & & & & \\
  LPNS$^\mathrm{s2}_\mathrm{YL04}$ & 0.1242 & 0.1627 & 0.2067 & 0.3270 & 0.4411 &
                                0.5434 & 0.7089 & 0.8276 & 0.9055 & 0.9594\\
  LPNS$^\mathrm{s1}_\mathrm{YL04}$ & 0.1046 & 0.1320 & 0.1737 & 0.2968 & 0.4172 &
                                0.5240 & 0.7008 & 0.8257 & 0.9113 & 0.9617\\
  HNS$^\mathrm{s2}$           & 0.1011 & 0.1473 & 0.1917 & 0.3128 & 0.4280 &
                                0.5305 & 0.6963 & 0.8176 & 0.9073 & 0.9750\\
  HNS$^\mathrm{s1}$           & 0.0684 & 0.1095 & 0.1559 & 0.2735 & 0.3905 &
                                0.4962 & 0.6665 & 0.7921 & 0.8862 & 0.9606\\
  CNS                         & 0.0507 & 0.0807 & 0.1255 & 0.2469 & 0.3699 &
                                0.4802 & 0.6569 & 0.7863 & 0.8812 & 0.9543\\
   & & & & & &  & & & & \\
  \hline
  \end{tabular}
\end{table*}
The speed of sound $v_\mathrm{s}$ in units of the speed of light $c$:
\begin{equation}
\label{3.5}
v_\mathrm{s} = \sqrt{\frac{\mathrm{d}P}{\mathrm{d}\varepsilon}}
             = \sqrt{\gamma \frac{P}{\varepsilon + P}},
\end{equation}
is tabulated in Table~\ref{soundspeed} for different EOS's.  The speed
of sound $v_\mathrm{s}$ increases with density up to nearly the speed
of light, in the most massive stars of a sequence.  Nevertheless,
$v_\mathrm{s}$ is always smaller than the speed of light and all EOS's
used in this paper are causal.  The speed of sound increases with
temperature and lepton number at fixed density in contrast to the
results of \cite{GHZ98}. This is probably caused by the
smaller temperature dependence of the EOS of 
\cite{LS91}, which was discussed before in this section.

\section{Structure of rotating and non-rotating protoneutron stars} \label{sec4}

The structure of rotating PNS's and NS's is governed by the Einstein
equations in stationary, axisymmetric, and asymptotic flat
space-time. Under these special conditions to the space-time symmetry
the ten Einstein equations reduce to four non-trivial equations which
are elliptic in quasi-isotropic coordinates \cite[]{BGSM93}. The
four non trivial Einstein equations together with the energy-momentum
conservation are solved via a finite difference scheme
\cite[]{Schaab98}. We follow \cite{BGSM93} in compactifying
the outer space to a finite region by using the transformation $r
\rightarrow 1/r$. The boundary condition of approximate flatness can
then be exactly fulfilled. The neutron star model is uniquely
determined by fixing one of the parameters: central density,
gravitational mass, or baryon number, as well as one of the
parameters: angular velocity, angular momentum, or stability parameter
$\beta=E_\mathrm{kin}/|E_\mathrm{grav}|$. The models of maximum mass 
and/or maximum rotation velocity can also be calculated.

\subsection{Protoneutron star and neutron star sequences} \label{ssec:pns.seq}

\begin{figure}
  \resizebox{\hsize}{!}{\rotatebox{-90}{\includegraphics{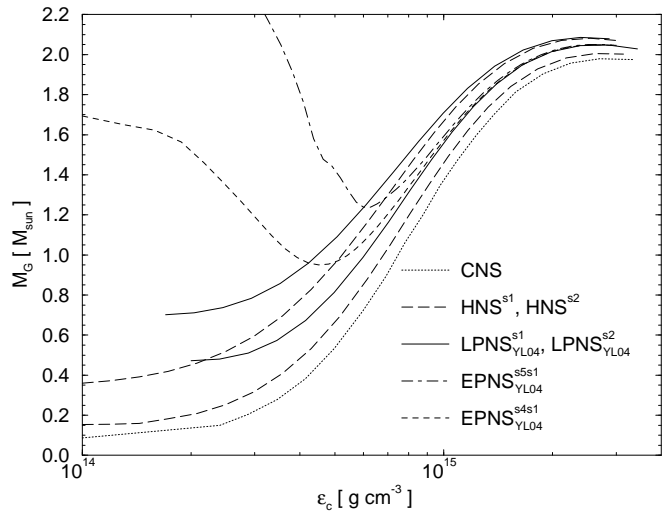}}}
  \caption[]{The gravitational mass versus central energy-density 
             of non-rotating NS's and PNS's.
             The lower long dashed line corresponds to the 
             HNS$^\mathrm{s1}$ EOS and the upper long dashed 
             line to the HNS$^\mathrm{s2}$ EOS.
             The lower solid line corresponds to the  
             LPNS$^\mathrm{s1}_\mathrm{YL04}$ EOS and the upper solid
             line to the LPNS$^\mathrm{s2}_\mathrm{YL04}$ EOS.}
  \label{mg_ec_nrot}
\end{figure}
Figure~\ref{mg_ec_nrot} shows the gravitational mass as function of the
central energy-density for the non-rotating PNS and NS models.  
Only models whose gravitational masses increase with central 
energy-density are stable against axisymmetric perturbations.
Whereas the maximum mass differs only by $\sim 5\,\%$ between the 
various star models, the minimum mass of the EPNS models,
$M_{\rm min}\sim 0.9-1.2M_{\sun}$, is much larger than the minimum mass
of the CNS models, $M_{\rm min}\sim 0.1M_{\sun}$.

\begin{figure}
  \resizebox{\hsize}{!}{\rotatebox{-90}{\includegraphics{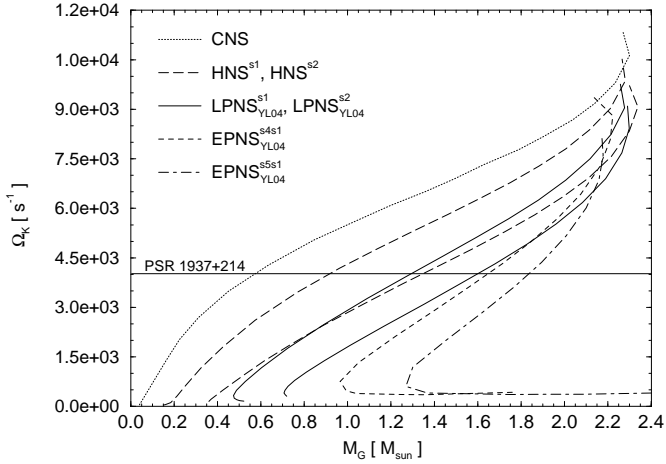}}}
  \caption[]{Kepler frequency versus gravitational mass of PNS's and NS's 
             for different EOS's. The upper long dashed line corresponds 
             to the HNS$^\mathrm{s1}$ EOS and the lower long dashed 
             line to the HNS$^\mathrm{s2}$ EOS.
             The upper solid line corresponds to the  
             LPNS$^\mathrm{s1}_\mathrm{YL04}$ EOS and the lower solid
             line to the LPNS$^\mathrm{s2}_\mathrm{YL04}$ EOS.}
  \label{omegaK}
\end{figure}
Figure \ref{omegaK} shows the Kepler frequency, i.e. the frequency at
which mass shedding sets in, as function of the gravitational mass.
Also shown is the rotational frequency, $\Omega=4033$~s$^{-1}$, of the
fastest pulsar known, PSR 1937+214 \cite[]{Bac82}. The
gravitational mass of this pulsar is unfortunately unknown but is
typically assumed to be in the range 1.0--2.0\,$M_{\sun}$. Since the
Kepler frequency of the CNS models is larger than the rotational
frequency of PSR 1937+214 for this mass range, this model is
consistent with the observation. 
 
\begin{figure}
  \resizebox{\hsize}{!}{\rotatebox{-90}{\includegraphics{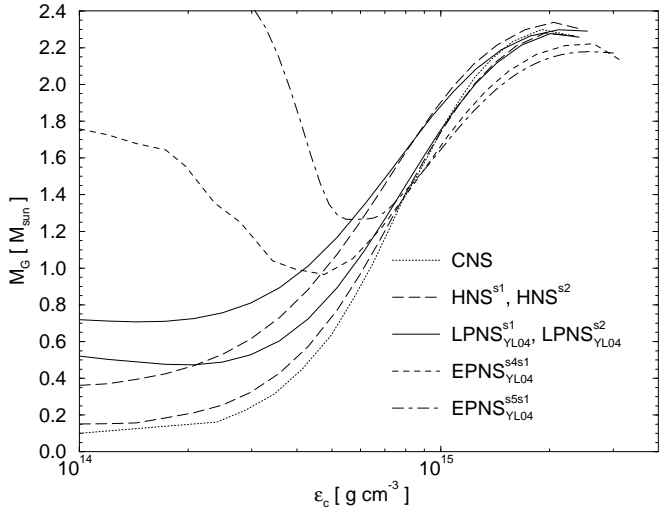}}}
  \caption[]{The gravitational mass versus central energy-density 
             of NS's and PNS's rotating at their Kepler frequency.
             The lower long dashed line corresponds to the 
             HNS$^\mathrm{s1}$ EOS and the upper long dashed 
             line to the HNS$^\mathrm{s2}$ EOS.
             The lower solid line corresponds to the  
             LPNS$^\mathrm{s1}_\mathrm{YL04}$ EOS and the upper solid
             line to the LPNS$^\mathrm{s2}_\mathrm{YL04}$ EOS.}
  \label{mg_ec_rot}
\end{figure}
Figure~\ref{mg_ec_rot} shows the sequences of stars rotating with
Kepler frequency.  As expected the masses are now larger at fixed
central energy density, but the maximum mass is reached for lower
central energy densities (see also Table~\ref{max_bar_mass}).  The
mass increase due to rotation is smaller for PNS's than for CNS's,
since the radii of PNS's are larger and thus the Kepler frequencies are
smaller.

\begin{figure}
  \resizebox{\hsize}{!}{\rotatebox{-90}{\includegraphics{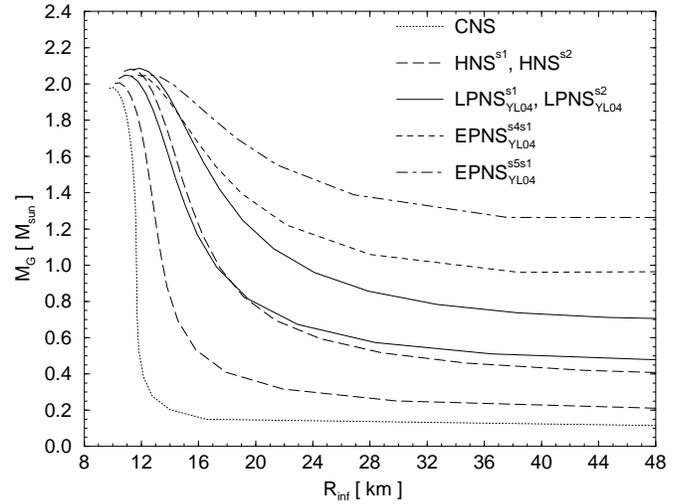}}}
  \caption[]{The gravitational mass versus stellar radius (as measured by
             an observer located at infinity) of non-rotating NS's and PNS's.
             The abbreviations for the different EOS's are 
             described in Table~\ref{EOSs}.
             The lower long dashed line corresponds to the 
             HNS$^\mathrm{s1}$ EOS and the upper long dashed 
             line to the HNS$^\mathrm{s2}$ EOS.
             The lower solid line corresponds to the  
             LPNS$^\mathrm{s1}_\mathrm{YL04}$ EOS and the upper solid
             line to the LPNS$^\mathrm{s2}_\mathrm{YL04}$ EOS.}
  \label{mg_r_nrot}
\end{figure}
\begin{figure}
  \resizebox{\hsize}{!}{\rotatebox{-90}{\includegraphics{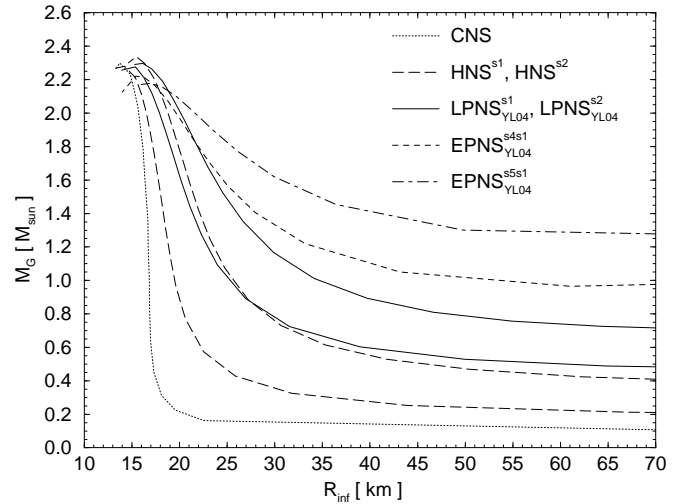}}}
  \caption[]{The gravitational mass versus equatorial radius (as measured by
             an observer located at infinity) of NS's and PNS's
             rotating at their Kepler frequency.
             The lower long dashed line corresponds to the 
             HNS$^\mathrm{s1}$ EOS and the upper long dashed 
             line to the HNS$^\mathrm{s2}$ EOS.
             The lower solid line corresponds to the  
             LPNS$^\mathrm{s1}_\mathrm{YL04}$ EOS and the upper solid
             line to the LPNS$^\mathrm{s2}_\mathrm{YL04}$ EOS.}
  \label{mg_r_rot}
\end{figure}
We also show the mass-radius relation for PNS's in
Figs.~\ref{mg_r_nrot} and \ref{mg_r_rot}\footnote{The equatorial
radius $R_\mathrm{inf}$ is the circumferential radius as measured in
infinity \cite[see, e.g.,][]{Schaab98}}.  In comparison with CNS's,
where the radius only slightly changes in the relevant mass region
around $M_\mathrm{G} \sim 1.4\,M_{\sun}$, one obtains for PNS's a
much stronger increase of the radius with decreasing mass, which is caused
by the stiffer EOS's for the PNS's.

\begin{figure}
  \resizebox{\hsize}{!}{{\includegraphics{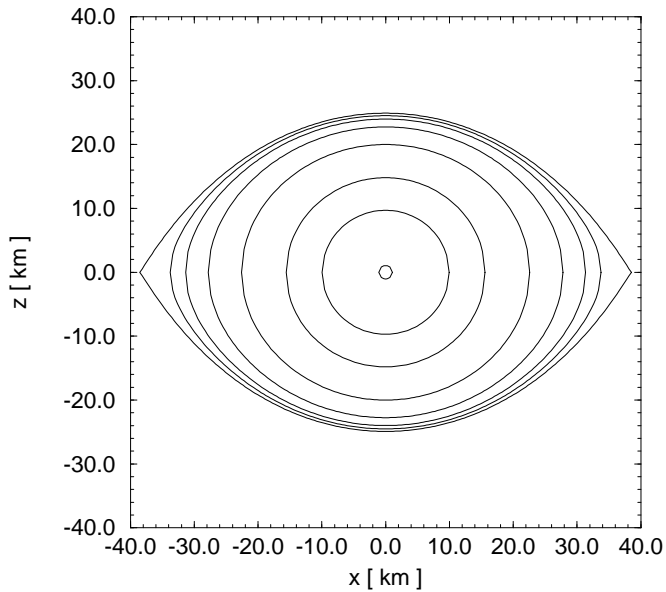}}}
  \caption[]{Iso-energy-density surfaces for a
  EPNS$^\mathrm{s5s1}_\mathrm{YL04}$ model with a baryonic mass of
  $1.5 M_{\sun}$ rotating at its Kepler frequency.  The surfaces
  correspond to energy-densities (beginning at the centre),
  $\varepsilon=0.99,0.1,0.01,10^{-3},10^{-4},10^{-5},10^{-6} 
  \varepsilon_\mathrm{c}$,
  where $\varepsilon_\mathrm{c}=8.07\times10^{14}$~g~cm$^{-3}$ denotes 
  the central energy-density.}  \label{edenss5}
\end{figure}
\begin{figure}
  \resizebox{\hsize}{!}{{\includegraphics{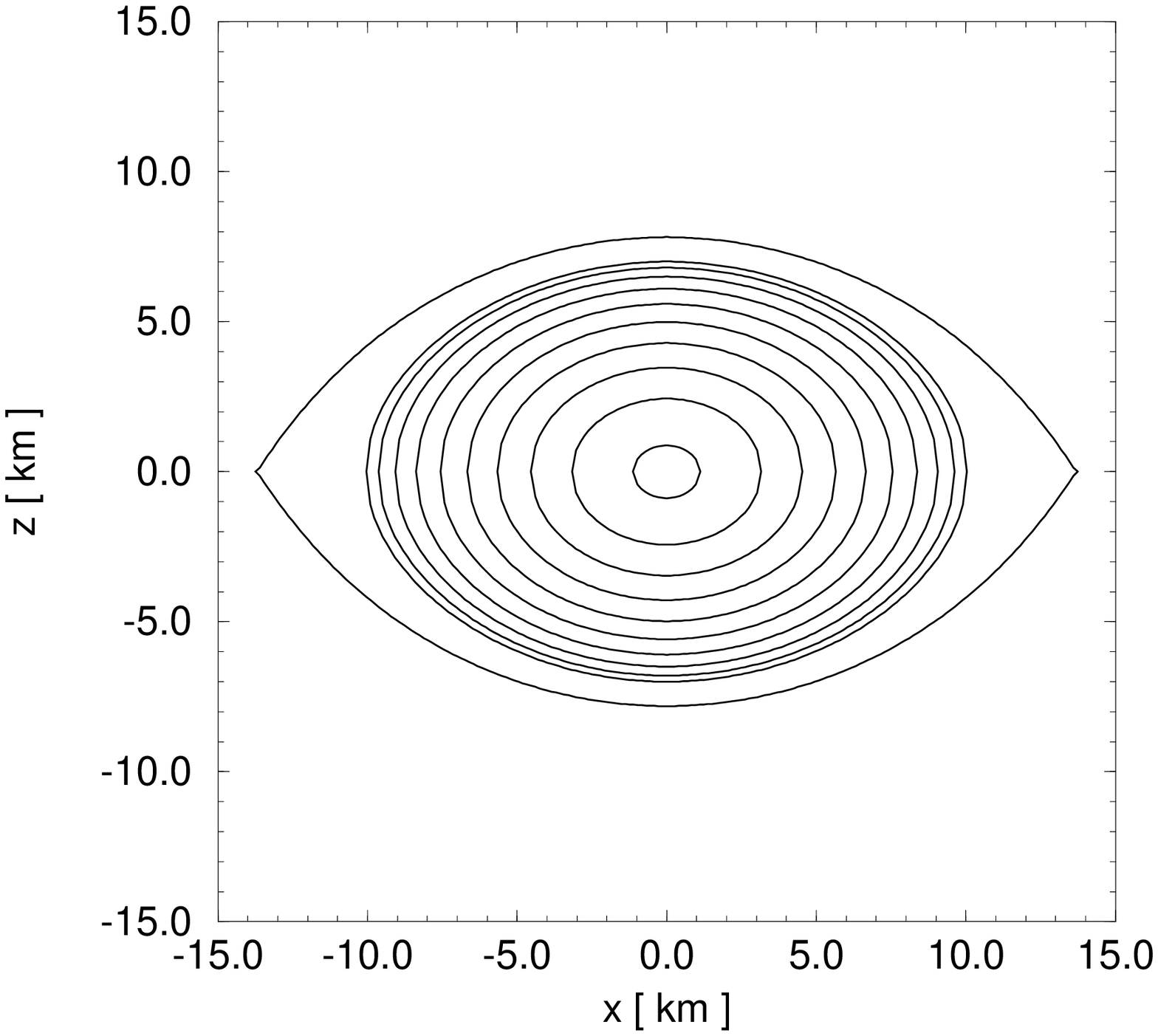}}}
  \caption[]{Iso-energy-density surfaces for a
  CNS model with a baryonic mass of
  $1.5 M_{\sun}$ rotating at its Kepler frequency.  The surfaces
  correspond to energy-densities (beginning at the centre),
  $\varepsilon=0.99,0.9,0.8,0.7,0.6,0.5,0.4,0.3,0.2,0.1 
  \varepsilon_\mathrm{c}$,
  where $\varepsilon_\mathrm{c}=8.11\times10^{14}$~g~cm$^{-3}$ denotes 
  the central energy-density.}  \label{edensT0}
\end{figure}
To demonstrate the differences of the internal structure of an EPNS
and a CNS, we show in Figs.~\ref{edenss5} and \ref{edensT0} the
iso-energy-density surfaces for models with $M_{\rm B}=1.5M_{\sun}$.  It
turns out that EPNS's contain mainly matter of densities below nuclear
matter density (Fig.~\ref{edenss5}).  Contrary to this, matter in a
CNS is dominated by matter of densities around and above nuclear
matter density (Fig.~\ref{edensT0}).

Another important point concerns the stability of the star models
against secular or dynamical instabilities. A certain configuration is
dynamically stable against axisymmetric perturbations if the
gravitational mass is minimum with respect to variations at fixed
baryon number and angular momentum. Along a star sequence with fixed
angular momentum, this is the case if the gravitational mass increases
with the central energy density \cite[see][p. 151]{ST83}. Please
note that the angular momentum is not fixed in the sequences of stars
rotating with Kepler frequency. The maximum mass configuration differs
therefore slightly from the marginally stable configuration. A fully
general relativistic analysis of dynamical and secular stability
against non-axisymmetric perturbations is extremely difficult and has
been performed only by means of approximations and/or special
assumptions in literature. It turned out however that the
configurations can be classified by a stability parameter defined by
the ratio of the kinetical energy and the absolute value of the
gravitational energy, $\beta = E_\mathrm{kin}/|E_\mathrm{grav}|$. For
values, $\beta\gtrsim 0.26$, the models are probably dynamically
unstable against the bar mode ($m=2$), whereas they become already
secularly unstable for $\beta\gtrsim 0.14$ \cite[see,
e.g.,][]{Dur75,Man85,IFD85,BFG96}. Higher modes,
$m\gtrsim 5$, of perturbations are suppressed by the finite viscosity
of neutron star matter. It is however uncertain whether modes with
$m=3$ or 4 are also suppressed. If this is not the case, secular
instability sets already in above a critical value, $\beta\gtrsim
0.08$ \cite[]{FIP86}.

\begin{figure}
  \resizebox{\hsize}{!}{\rotatebox{-90}{\includegraphics{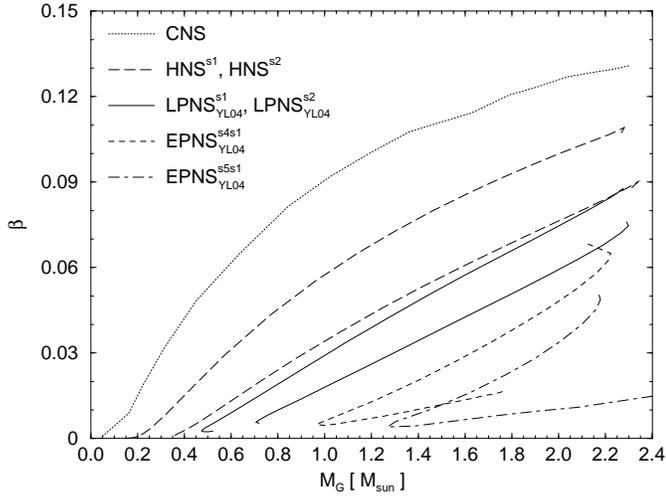}}}
  \caption[]{Stability parameter $\beta = E_\mathrm{kin}/|E_\mathrm{grav}|$ 
             versus gravitational mass
             of PNS's and NS's rotating at their Kepler frequency
             for different EOS's. The upper long dashed line corresponds 
             to the HNS$^\mathrm{s1}$ EOS and the lower long dashed 
             line to the HNS$^\mathrm{s2}$ EOS.
             The upper solid line corresponds to the  
             LPNS$^\mathrm{s1}_\mathrm{YL04}$ EOS and the lower solid
             line to the LPNS$^\mathrm{s2}_\mathrm{YL04}$ EOS.}
  \label{e_rot_e_grav}
\end{figure}
Figure~\ref{e_rot_e_grav} shows the value of the stability parameter
for stars rotating with their Kepler frequency. Since the timescale
for the growing of the secular instabilities is larger than the
evolution timescale of PNS's and HNS's, only the dynamical instability
is important for PNS's and HNS's. The critical value for the dynamical
instability, $\beta\gtrsim 0.26$, is reached by none of the
models. Depending on the internal structure of the CNS's, the CNS's
models rotating with Kepler frequency may become secular unstable
against modes with $m=3$ or $m=4$ for star masses $M\gtrsim 0.8
M_{\sun}$.

\subsection{Evolution of a non-rotating protoneutron star} \label{ssec:evol.seq}

\begin{figure}
  \resizebox{\hsize}{!}{\rotatebox{-90}{\includegraphics{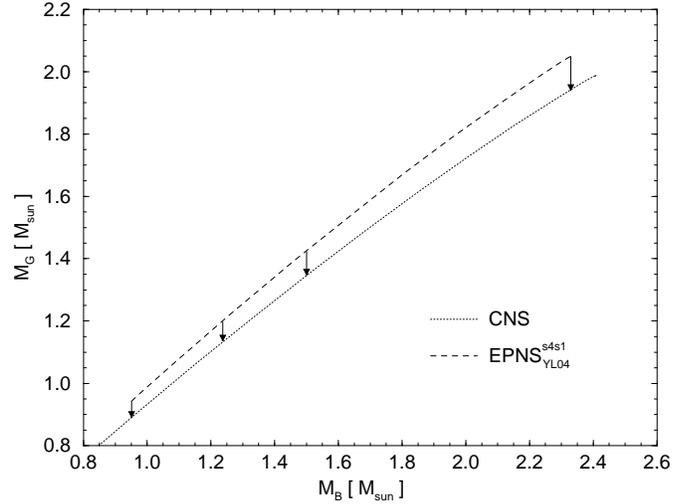}}}
  \caption[]{Gravitational mass versus baryonic mass for the 
             non-rotating EPNS$^\mathrm{s4s1}_\mathrm{YL04}$ EOS and 
             the non-rotating CNS EOS.
             The EPNS$^\mathrm{s5s1}_\mathrm{YL04}$ EOS lies 
             exactly on the EPNS$^\mathrm{s4s1}_\mathrm{YL04}$ 
             line and is not shown by that reason.
             The arrows show the evolution (from left to right)
             for the minimum mass 
             EPNS$^\mathrm{s4s1}_\mathrm{YL04}$, the minimum mass 
             EPNS$^\mathrm{s5s1}_\mathrm{YL04}$, the 
             $M_\mathrm{B} = 1.5~M_{\sun}$, and the maximum mass
             configurations.}
  \label{mg_mb}
\end{figure}

\begin{table*}
  \caption[]{Properties of non-rotating and with Kepler frequency rotating 
             PNS's and NS's,
             for a fixed baryonic mass $M_\mathrm{B} = 1.5\,M_{\sun}$.
             The EOS's are 
             summarized in Table~\ref{EOSs}. 
             The entries are: gravitational mass, 
             $M_\mathrm{G}$;
             baryonic mass, $M_\mathrm{B}$;
             circumferential radius (as measured in infinity), 
             $R_\mathrm{inf}$;
             central baryon number density, $n_\mathrm{c}$;
             central temperature, $T_\mathrm{c}$;
             Kepler frequency, $\Omega_\mathrm{K}$;
             angular momentum, $J$
             ($M_{\sun}$~km $\hat =\,5.966\times10^{48}$ g~cm$^{2}$~s$^{-1}$).}
  \label{neutrinosphere}
  \begin{tabular}{ l c c c c @{~~~~~~~~~~} c c c c c }
  \hline
   & & & & & & & & & \\
   & \multicolumn{4}{c}{$\Omega = 0$} &  
   \multicolumn{5}{c}{$\Omega = \Omega_\mathrm{K}$}   \\
   & & & & & & & & & \\
   \cline{2-10}
   & & & & & & & & & \\
  EOS  & $M_\mathrm{G}$ & 
         $R_\mathrm{inf}$ & $n_\mathrm{c}$ & $T_\mathrm{c}$ &
         $M_\mathrm{G}$ & 
         $R_\mathrm{inf}$ & $n_\mathrm{c}$ & $\Omega_\mathrm{K}$ 
         & $J$ \\
       & [$M_{\sun}$]  & [km]  & [fm$^{-3}$] & [MeV] &
         [$M_{\sun}$]  & [km]   & [fm$^{-3}$] & [s$^{-1}$]  &
         [$M_{\sun}$ km] \\
   & & & & & & & & & \\
  \hline
   & & & & & & & & & \\
  EPNS$^\mathrm{s5s1}_\mathrm{YL04}$    & 1.425 & 25.66 & 0.440 & 20.8 &
                                          1.430 & 38.55 & 0.427 & 1879 & 0.634\\
  EPNS$^\mathrm{s4s1}_\mathrm{YL04}$    & 1.425 & 18.66 & 0.456 & 21.3 &
                                          1.433 & 27.46 & 0.436 & 3056 & 0.924\\
  LPNS$^\mathrm{s2}_\mathrm{YL04}$      & 1.431 & 17.38 & 0.385 & 39.5 &
                                          1.437 & 25.62 & 0.353 & 3410 & 1.234\\
  LPNS$^\mathrm{T06s2}_\mathrm{YL04}$   & 1.432 & 17.47 & 0.386 & 39.6 &
                                          1.437 & 25.77 & 0.355 & 3381 & 1.219\\
  LPNS$^\mathrm{s2}_\mathrm{YL04(64-63)}$ & 1.430 & 16.72 & 0.386 & 39.6 &
                                          1.438 & 24.68 & 0.348 & 3610 & 1.340\\
  LPNS$^\mathrm{s2}_\mathrm{YL04(64-22)}$ & 1.430 & 16.34 & 0.386 & 39.6 &
                                          1.438 & 24.13 & 0.344 & 3739 & 1.408\\
  LPNS$^\mathrm{s2}_\mathrm{YL04(64-62)}$ & 1.430 & 15.85 & 0.388 & 39.7 &
                                          1.439 & 23.36 & 0.341 & 3930 & 1.497\\
  LPNS$^\mathrm{s2}_\mathrm{YL03}$      & 1.414 & 16.01 & 0.400 & 42.9 &
                                          1.423 & 23.37 & 0.361 & 3897 & 1.362\\
  LPNS$^\mathrm{s1}_\mathrm{YL04}$      & 1.411 & 14.70 & 0.459 & 21.4 &
                                          1.423 & 21.26 & 0.417 & 4482 & 1.389\\
  LPNS$^\mathrm{s1}_\mathrm{YL03}$      & 1.391 & 13.59 & 0.475 & 22.8 &
                                          1.406 & 19.55 & 0.424 & 5076 & 1.552\\
  LPNS$^\mathrm{s0}_\mathrm{YL04}$      & 1.410 & 12.39 & 0.519 & 0 &
                                          1.429 & 17.73 & 0.450 & 5949 & 1.800\\
  HNS$^\mathrm{s2}$                     & 1.391 & 15.22 & 0.412 & 49.7 &
                                          1.403 & 22.13 & 0.367 & 4204 & 1.431\\
  HNS$^\mathrm{T06s2}$                  & 1.392 & 15.29 & 0.412 & 49.7 &
                                          1.404 & 22.24 & 0.369 & 4171 & 1.413\\
  HNS$^\mathrm{T0s2}$                   & 1.385 & 13.53 & 0.404 & 49.0 &
                                          1.407 & 19.86 & 0.324 & 5075 & 2.072\\
  HNS$^\mathrm{s1}$                     & 1.362 & 12.78 & 0.501 & 26.8 &
                                          1.382 & 18.36 & 0.437 & 5562 & 1.669\\
  HNS$^\mathrm{T03s1}$                  & 1.363 & 12.80 & 0.502 & 26.9 &
                                          1.382 & 18.39 & 0.438 & 5550 & 1.662\\
  HNS$^\mathrm{T0s1}$                   & 1.362 & 12.18 & 0.504 & 27.0 &
                                          1.387 & 17.59 & 0.418 & 6023 & 1.957\\
  CNS                                   & 1.346 & 11.45 & 0.551 & 0 &
                                          1.374 & 16.36 & 0.459 & 6689 & 1.953\\
   & & & & & & & & & \\
  \hline
  \end{tabular}
\end{table*}
The evolution of a PNS to a CNS can be followed by means of several
``snapshots'' taken at different times after core bounce (see
Sect.~\ref{sec2}).  The evolution path is determined by fixing the
baryonic mass and the angular momentum if accretion of matter and loss of
angular momentum is neglected. First, we study the evolution of a
non-rotating PNS with $M_{\rm B}=1.5M_{\sun}$ (see Table
\ref{neutrinosphere} and Fig. \ref{mg_mb}). 

The first snapshot corresponds to approximately 50 - 100 ms when the
envelope is characterized by a high entropy per baryon and high lepton
number (see curves labeled EPNS$^\mathrm{s5s1}_\mathrm{YL04}$ and
EPNS$^\mathrm{s4s1}_\mathrm{YL04}$). After 0.5 - 1 seconds the PNS
reached our LPNS stage, which is characterized by an approximately
constant entropy per baryon, $s\sim 2$, throughout the star (model
LPNS$^\mathrm{s2}_\mathrm{YL04}$). Due to the higher entropy per
baryon in the core of the PNS, the central density and thus the
gravitational binding energy decreases. On the other hand the entropy
per baryon decreases in the envelope and therefore the radius
decreases, too. For a lower value of the entropy per baryon and/or a
lower value for the lepton number, the gravitational binding energy
increases compared to the EPNS stage (models
LPNS$^\mathrm{s2}_\mathrm{YL03}$, LPNS$^\mathrm{s1}_\mathrm{YL04}$,
LPNS$^\mathrm{s1}_\mathrm{YL03}$).  After about 10 - 30 seconds the
neutrinos escape from the star and its EOS softens (HNS models). The
gravitational binding energy thus increases by roughly 3\,\%. Finally,
the CNS model is even more compressed. 

Due to a smaller increase of the pressure with increasing temperature, 
\cite{BL86} and \cite{PRPLM98} obtain a monotonous
increase of the central density during the evolution of the PNS to the CNS. 
In contrast, our results, and also those of \cite{KJ95}, show a more
complex behaviour of the central density \cite[see discussion in the
paper of][]{PRPLM98}.
This difference has mainly two reasons: 
Firstly, \cite{PRPLM98} use an entropy profile obtained in a supernova 
collapse simulation 
of a 1.08~$M_{\sun}$ NS, which leads to a higher central entropy per 
baryon, $s \sim 1.5$, at the EPNS stage and an increase to $s \sim 2$
at the LPNS stage. We use the calculations of \cite{BHF95} who
simulated  the supernova collapse of a NS with $M \sim 1.5~M_{\sun}$. 
They obtained a smaller starting central entropy per baryon, 
$s \sim 1$, and an increase to $s \sim 2$ at the LPNS stage.
The second reason is the fact, that they use an approximation for 
the temperature influence on the pressure \cite[derived by][]{Pra97} 
of the nucleons:
\begin{equation}
\label{4.1}
s = \frac{\pi^2}{2}\,T \sum_{\tau = n, p} 
                      \frac{Y_\mathrm{\tau}}{T_\mathrm{F, \tau}} ,
\end{equation}
\begin{equation}
\label{4.1a}
\frac{P_\mathrm{th}}{P_0} = \frac{5 s}{3 \pi^2} \frac{\sum\nolimits_\tau \frac{Y_\tau}{T_\mathrm{F, \tau}} \left( 1 - \frac{3}{2} \frac{\mathrm{d~ln} m^*_\tau}{\mathrm{d~ln} n_\tau} \right)}{\left(\sum\nolimits_\tau \frac{Y_\tau}{T_{F, \tau}}\right)^2\left(\sum\nolimits_\tau Y_\tau T_{F, \tau}\right)}\left( 1 + \frac{P_\mathrm{pot}}{P_\mathrm{kin}}\right)^{-1} ,
\end{equation}
where $T_\mathrm{F, \tau}$ denotes the Fermi temperature of the quasi particles 
($T_\mathrm{F, \tau} = p_\mathrm{F, \tau}^2/2m^*_\tau$), $P_\mathrm{th}$, the 
thermal pressure and 
$P_0 = P_\mathrm{kin} + P_\mathrm{pot}$, the pressure at zero temperature 
as sum of kinetic and potential pressure.
This approximation holds under the assumption that in the dense parts 
of the PNS the temperature is small in comparison with 
the Fermi temperature, i.e. $T/T_\mathrm{F} \ll 1$. However in our case 
these ratio reaches $T/T_\mathrm{F} \approx 0.45$ 
($T/T_\mathrm{F} \approx 0.6$) at $n = n_{0}$ and $T/T_\mathrm{F} \approx 0.29$
($T/T_\mathrm{F} \approx 0.44$) at $n = 1$ fm$^{-3}$ for neutrons (protons) in 
the LPNS$^\mathrm{s2}_\mathrm{YL04}$ model (see also the Fermi-Dirac 
distribution function in Fig.~\ref{fn} of Appendix A and 
\cite{TNH94} for this purpose, in which it can be seen that more than
$10\,\%$ of the matter is non-degenerate).
Therefore this approximation underestimates the pressure increase due to 
thermal effects, e.g. the thermal pressure $P_\mathrm{th}$ is exact 
(approximative) 2.37~MeVfm$^{-3}$ (1.26~MeVfm$^{-3}$) at $n = n_0$ and 
54.62~MeVfm$^{-3}$ (28.83~MeVfm$^{-3}$) at $n = 1$~fm$^{-3}$ for the 
nucleons of the LPNS$^\mathrm{s2}_\mathrm{YL04}$ case of our model. Hence 
in our opinion, higher order terms should be included in the treatment.
Since the temperature dependence of the GM3 model of \cite{PRPLM98} is 
smaller than in our case, the deviations between the exact solution and 
the approximation may be small.

\subsection{Maximum rotational frequency of a neutron star}

\begin{table*}
  \caption[]{Evolution of 1.5 $M_{\sun}$ star at constant angular
  momentum, $J=0.634$~$M_{\sun}$\,km and
  $J=0.924$~$M_{\sun}$\,km for the sequences
  starting with the model EPNS$^\mathrm{s5s1}_\mathrm{YL04}$ and
  EPNS$^\mathrm{s4s1}_\mathrm{YL04}$, respectively.}
  \label{evolv15}
  \begin{tabular}{ l c c c c c  @{~~~~~~~~~~} c c c c c   }
  \hline
   & & & & & & & & & & \\
   & \multicolumn{5}{c}{EPNS$^\mathrm{s5s1}_\mathrm{YL04}$} &  
   \multicolumn{5}{c}{EPNS$^\mathrm{s4s1}_\mathrm{YL04}$}   \\
   & & & & & & & & & &  \\
   \cline{2-11}
   & & & & & & & & & & \\
  EOS  & $M_\mathrm{G}$ & $R_\mathrm{inf}$ & $n_\mathrm{c}$ & 
         $T_\mathrm{c}$ & $\Omega$ 
       & $M_\mathrm{G}$ & $R_\mathrm{inf}$ & $n_\mathrm{c}$ & 
         $T_\mathrm{c}$ & $\Omega$    \\
       & [$M_{\sun}$]  & [km]  & [fm$^{-3}$] & [MeV] & [s$^{-1}$] 
       & [$M_{\sun}$]  & [km]  & [fm$^{-3}$] & [MeV] & [s$^{-1}$] \\
   & & & & & & & & & & \\
  \hline
   & & & & & & & & & & \\
  EPNS                                  & 1.430 & 38.55 & 0.427 & 20.4 & 1879 &
                                          1.433 & 27.46 & 0.436 & 20.7 & 3056\\
  LPNS$^\mathrm{s2}_\mathrm{YL04}$      & 1.349 & 14.83 & 0.350 & 37.0 & 2091 &
                                          1.341 & 15.40 & 0.336 & 36.0 & 2961 \\
  LPNS$^\mathrm{s2}_\mathrm{YL03}$      & 1.363 & 14.27 & 0.374 & 40.9 & 2171 &
                                          1.361 & 14.74 & 0.362 & 39.9 & 3073 \\
  LPNS$^\mathrm{s1}_\mathrm{YL04}$      & 1.391 & 13.17 & 0.442 & 20.9 & 2377 & 
                                          1.391 & 13.54 & 0.430 & 20.5 & 3390 \\
  LPNS$^\mathrm{s1}_\mathrm{YL03}$      & 1.385 & 12.76 & 0.462 & 22.4 & 2474 &
                                          1.387 & 13.09 & 0.451 & 22.0 & 3525 \\
  HNS$^\mathrm{s2}$                     & 1.349 & 14.06 & 0.387 & 46.5 & 2246 &
                                          1.348 & 14.53 & 0.375 & 46.3 & 3178 \\
  HNS$^\mathrm{s1}$                     & 1.362 & 12.49 & 0.489 & 26.3 & 2609 &
                                          1.365 & 12.83 & 0.478 & 25.9 & 3709 \\
  CNS                                   & 1.348 & 11.83 & 0.535 & 0 & 2833 &
                                          1.351 & 12.14 & 0.523 & 0 & 4031 \\
   & & & & & & & & & & \\
  \hline
  \end{tabular}
\end{table*}
We follow now the evolution of an EPNS with $M_{\rm B}=1.5M_{\sun}$,
which rotates with Kepler frequency at its EPNS state. During its
evolution, the angular momentum is assumed to be conserved (see Table
\ref{evolv15}). As in the non-rotating case, the star becomes more and
more compressed during its evolution. With only a few exceptions, the
central baryon density increases, whereas the gravitational mass and
the circumferential radius decreases. It is obvious that this trend
has to be counterbalanced by an increasing angular velocity in order
to keep the angular momentum constant. Compared to the EPNS state, the
angular velocity in the CNS state is increased by 51\,\% or 32\,\% for
the sequences starting with the model
EPNS$^\mathrm{s5s1}_\mathrm{YL04}$ or
EPNS$^\mathrm{s4s1}_\mathrm{YL04}$, respectively. Nevertheless, the
angular velocity in the CNS state reaches at most 60\,\% of the Kepler
frequency. 

\begin{figure}
  \resizebox{\hsize}{!}{\rotatebox{-90}{\includegraphics{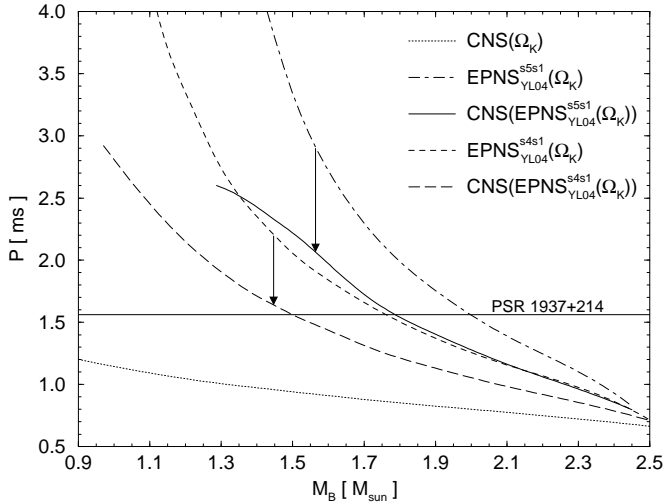}}}
  \caption[]{Minimum rotational period $P$ versus baryonic mass of the
  EPNS's and CNS's.  The figure shows the spin up during the evolution
  of the EPNS's to CNS's.  The dot-dashed (short dashed) line shows
  the EPNS$^\mathrm{s5s1}_\mathrm{YL04}$
  (EPNS$^\mathrm{s4s1}_\mathrm{YL04}$) model rotating at its Kepler
  frequency. The solid (long dashed) line shows the corresponding CNS
  model obtained by cooling at constant angular momentum. The dotted
  line shows CNS's rotating at their Kepler frequency for comparison.
  The arrows represent the evolution from the EPNS stage to the CNS
  stage.}  \label{maxrot}
\end{figure}
In this respect, the minimum rotational period is determined by the
Kepler rotating EPNS model. Figure \ref{maxrot} shows the the general
evolution of the rotational period for the two EPNS models. For a CNS
with a typical baryonic mass of $1.5 M_{\sun}$ one obtains a minimum
rotational period, $P$, between $1.56$ and $2.22$~ms. 
Recently, \cite{GHZ97, GHZ98} found similar results for the minimum 
rotational period, which confirms our calculations.
The period of
the fastest known pulsar PSR 1937+214, $P=1.56$~ms \cite[]{Bac82} is
at the lower limit of this range. Pulsars that rotate even faster
cannot be born with such small periods but have to be accelerated
after their formation, as long as a typical baryonic mass, $M_{\rm
B}=1.5 M_{\sun}$ is assumed. \cite{And98} has recently shown
that not too cold NS's rotating with $\Omega\gtrsim 0.1\Omega_{\rm K}$
are unstable against r-modes. This means that
the minimum rotational period for a young NS is even higher, 
$P_{\rm min}\sim 10$~ms.
Both the results by \cite{And98} and the results presented here
strengthen the view of millisecond pulsars as being recycled by
accretion \cite[]{Lor96}.

\begin{figure}
  \resizebox{\hsize}{!}{\rotatebox{-90}{\includegraphics{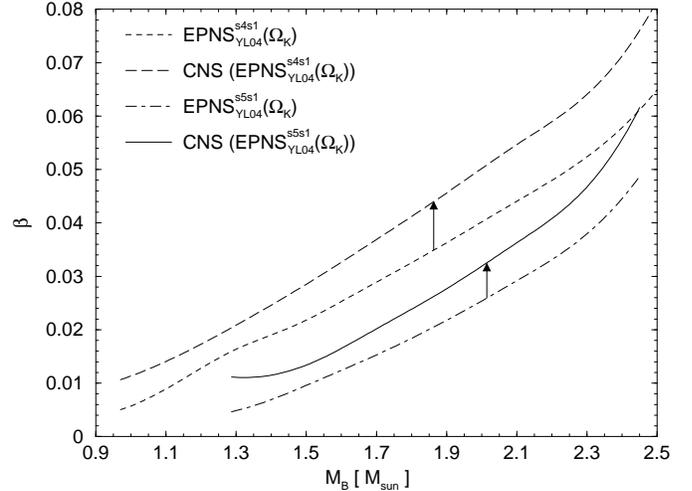}}}
  \caption[]{Variation of the stability parameter
  $\beta=E_\mathrm{kin}/|E_\mathrm{grav}|$ during the evolution from
  EPNS's to CNS's at constant angular momentum. The arrows represent 
  the evolution from the EPNS stage to the CNS stage.}  \label{beta_evolv}
\end{figure}
Figure~\ref{beta_evolv} shows the behaviour of $\beta$ during the
evolution from the EPNS to the CNS state. The stability parameter
increases during the evolution. Nevertheless, its maximum value,
$\beta\sim 0.082$, reached for the most massive star of the 
EPNS$^\mathrm{s4s1}_\mathrm{YL04}$ EOS, is only a little larger than
the critical value $\beta\sim 0.08$ for the onset of the secular
instabilities with $m=3,4$ (see Sect.\ \ref{ssec:pns.seq}).

\subsection{Maximum and minimum mass of a neutron star}

\begin{table*}
  \caption[]{Properties of the maximum gravitational mass configurations
             of non-rotating and with Kepler frequency rotating PNS's 
             and NS's.}
  \label{max_bar_mass}
  \begin{tabular}{ l c c c c c @{~~~~~~~~~~} c c c c c c }
  \hline
   & & & & & & & & &  & & \\
   & \multicolumn{5}{c}{$\Omega = 0$} &  
     \multicolumn{6}{c}{$\Omega = \Omega_\mathrm{K}$} \\
    & & & & & & & & & & & \\
    \cline{2-12}
   & & & & & & & & & & & \\
  EOS  & $M_\mathrm{G}$ & $M_\mathrm{B}$ & 
         $R_\mathrm{inf}$ & $n_\mathrm{c}$ & $ T_\mathrm{c}$ &
         $M_\mathrm{G}$ & $M_\mathrm{B}$ & 
         $R_\mathrm{inf}$ & $n_\mathrm{c}$ &
         $\Omega_\mathrm{K}$ & $J$\\
       & [$M_{\sun}$]  & [$M_{\sun}$]  & [km]   & [fm$^{-3}$]   & [MeV] &
         [$M_{\sun}$]  & [$M_{\sun}$]  & [km]  & [fm$^{-3}$] & [s$^{-1}$]
         & [$M_{\sun}$ km] \\
   & & & & & & & & & & & \\
  \hline
   & & & & & & & & & & & \\
  EPNS$^\mathrm{s5s1}_\mathrm{YL04}$      & 2.05 & 2.33 & 12.59 & 1.09 & 
                                          40.4 & 2.18 & 
                                          2.45 & 16.93 & 1.08 & 7609 & 3.06\\
  EPNS$^\mathrm{s4s1}_\mathrm{YL04}$      & 2.05 & 2.33 & 11.83 & 1.10 & 
                                          40.7 & 2.22 & 
                                          2.50 & 15.44 & 1.09 & 8803 & 3.69\\
  LPNS$^\mathrm{s2}_\mathrm{YL04}$        & 2.09 & 2.33 & 11.91 & 1.03 & 
                                          81.6 & 2.30 & 
                                          2.54 & 15.90 & 0.97 & 8602 & 4.29\\
  LPNS$^\mathrm{s2}_\mathrm{YL03}$        & 2.08 & 2.37 & 11.29 & 1.09 & 
                                          92.0 & 2.33 & 
                                          2.62 & 15.43 & 0.95 & 9067 & 4.61\\
  LPNS$^\mathrm{s1}_\mathrm{YL04}$        & 2.05 & 2.33 & 10.82 & 1.14 & 
                                          41.8 & 2.28 & 
                                          2.56 & 15.38 & 0.93 & 9384 & 4.49\\
  LPNS$^\mathrm{s1}_\mathrm{YL03}$        & 2.04 & 2.38 & 10.70 & 1.18 & 
                                          45.6 & 2.31 & 
                                          2.65 & 14.43 & 0.93 & 9701 & 4.94\\
  LPNS$^\mathrm{s0}_\mathrm{YL04}$        & 2.05 & 2.27 & 10.29 & 1.12 & 
                                          0 & 2.28 & 
                                          2.54 & 14.31 & 0.88 & 9956 & 4.98\\
  HNS$^\mathrm{s2}$                       & 2.08 & 2.41 & 11.24 & 1.12 & 
                                          120.6 & 2.34 & 
                                          2.68 & 15.47 & 0.92 & 9073 & 4.80\\
  HNS$^\mathrm{s1}$                       & 2.01 & 2.38 & 10.48 & 1.20 & 
                                          62.1 & 2.28 & 
                                          2.66 & 14.14 & 0.92 & 9864 & 5.03\\
  CNS                                     & 1.99 & 2.41 &  9.83 & 1.25 & 
                                          0 & 2.30 & 
                                          2.70 & 13.74 & 0.95 & 10631 & 5.64\\
   & & & & & & & & & & & \\
  \hline
  \end{tabular}
\end{table*}
Table \ref{max_bar_mass} shows, that the maximum baryonic mass of a
non-rotating (rotating) EPNS is $2.33 M_{\sun}$ ($2.45 - 2.5
M_{\sun}$), whereas the maximum baryonic mass of a CNS is 2.41
$M_{\sun}$ (2.7 $M_{\sun}$) (see also Fig.\ \ref{mg_mb}).  Since the
maximum baryonic mass of a LPNS and HNS is larger than the maximum
baryonic mass of a EPNS, too, it is obvious that a PNS based on our
nuclear EOS cannot collapse to a black hole during its
Kelvin-Helmholtz cooling phase if any further accretion is neglected
\cite[see][]{BJKST96}.  The situation changes if hyperons, meson
condensation, or a quark-hadron phase transition are included
\cite[see, e.g.,][]{BB94, PPT95, PCL95, Gle95}.
Another important point concerning the maximum mass of a CNS is 
due to the different maximum masses of EPNS's and CNS's rotating at 
their Kepler frequencies. The EPNS can support a baryonic mass of 
$2.45 - 2.5 M_{\sun}$, which is only slightly higher than the maximum 
baryonic mass of the non-rotating CNS ($2.41 M_{\sun}$), but 
$0.2 - 0.25 M_{\sun}$ less than the maximum baryonic mass of the CNS 
rotating at its Kepler frequency. Such supramassive CNS's enter unstable 
regions during their spin down evolution due to unstability against 
axisymmetric perturbations and may finally collapse to a black hole. 
For a discussion of these supramassive CNS's, see \cite{CST94} 
and \cite{SBGH94}.

In first approximation, the gravitational mass of non-rotating PNS's
and HNS's increases quadratically with the entropy per baryon, which is
taken constant throughout the star \cite[see][]{Pra97}:
\begin{equation}
\label{massincreas}
M_\mathrm{G}(s) = M_\mathrm{G}(T=0) \left( 1 + \lambda s^2 + \dots \right).
\end{equation}
We obtain for $\lambda$ approximate values $\sim 0.010 - 0.011$ 
($\sim 0.005$)
for the cases without (with) trapped lepton number (see
Table~\ref{max_bar_mass}). These values are in agreement with the
values derived by \cite{Pra97}.  In the investigation of \cite{GHZ97}
the gravitational mass shows the opposite behaviour: The gravitational
mass decreases slightly with increasing entropy per baryon. 
This is probably caused by the smaller temperature dependence of the 
EOS of \cite{LS91}, as it was pointed out in
Sect.~\ref{sec3}.  In the case of rotating stars the behaviour seems
to be more complex since the value of $\lambda$ is negative for
HNS$^\mathrm{s1}$ and LPNS$^\mathrm{s1}_\mathrm{YL04}$ and positive
for HNS$^\mathrm{s2}$ and LPNS$^\mathrm{s2}_\mathrm{YL04}$.  Another
interesting point is, that the maximum gravitational and baryonic mass
of the non-rotating PNS's are determined by the EOS used in the core
and do not depend on the EOS used in the envelope (compare the models
EPNS$^\mathrm{s4s1}_\mathrm{YL04}$,
EPNS$^\mathrm{s5s1}_\mathrm{YL04}$, and
LPNS$^\mathrm{s1}_\mathrm{YL04}$ in Table \ref{max_bar_mass} and
Figs.~\ref{mg_ec_nrot} and \ref{mg_r_nrot}). In the case of rotating
PNS's, the maximum gravitational mass of the EPNS is however smaller
compared to the corresponding LPNS model (see Table \ref{max_bar_mass}
and Figs.~\ref{mg_ec_rot} and \ref{mg_r_rot}). This behaviour is caused
by the higher Kepler frequency which can be supported by the LPNS
models (see Sect.\ \ref{ssec:pns.seq}).

\begin{table*}
  \caption[]{Evolution of minimum mass star for the non-rotating 
             EPNS$^\mathrm{s5s1}_\mathrm{YL04}$ EOS 
             ($M_\mathrm{B} = 1.238 M_{\sun}$)
             and the EPNS$^\mathrm{s4s1}_\mathrm{YL04}$ EOS  
             ($M_\mathrm{B} = 0.950 M_{\sun}$).}
  \label{evolvmin}
  \begin{tabular}{ l c c c c   @{~~~~~~~~~~} c c c c    }
  \hline
   & & & & & & &  \\
   & \multicolumn{4}{c}{EPNS$^\mathrm{s5s1}_\mathrm{YL04}$} &  
   \multicolumn{4}{c}{EPNS$^\mathrm{s4s1}_\mathrm{YL04}$}   \\
   & & & & & & &    \\
   \cline{2-9}
   & & & & & & &   \\
  EOS  & $M_\mathrm{G}$ & $R_\mathrm{inf}$ & $n_\mathrm{c}$ & 
         $T_\mathrm{c}$ 
       & $M_\mathrm{G}$ & $R_\mathrm{inf}$ & $n_\mathrm{c}$ & 
         $T_\mathrm{c}$    \\
       & [$M_{\sun}$]  & [km]  & [fm$^{-3}$] & [MeV] 
       & [$M_{\sun}$]  & [km]  & [fm$^{-3}$] & [MeV] \\
   & & & & & & &   \\
  \hline
   & & & & & & &  \\
  EPNS                                  & 1.207 & 45.61 & 0.340 & 17.3 & 
                                          0.942 & 44.49 & 0.281 & 15.1 \\
  LPNS$^\mathrm{s2}_\mathrm{YL04}$      & 1.199 & 19.77 & 0.308 & 33.8 & 
                                          0.926 & 25.34 & 0.219 & 26.7 \\
  LPNS$^\mathrm{s2}_\mathrm{YL03}$      & 1.186 & 17.49 & 0.327 & 37.1 &
                                          0.922 & 20.38 & 0.250 & 30.7 \\
  LPNS$^\mathrm{s1}_\mathrm{YL04}$      & 1.185 & 15.83 & 0.384 & 18.9 &  
                                          0.926 & 18.05 & 0.307 & 16.1 \\
  LPNS$^\mathrm{s1}_\mathrm{YL03}$      & 1.168 & 14.24 & 0.404 & 20.3 & 
                                          0.915 & 15.31 & 0.331 & 17.6 \\
  HNS$^\mathrm{s2}$                     & 1.168 & 16.28 & 0.341 & 43.2 & 
                                          0.911 & 18.26 & 0.267 & 36.5 \\
  HNS$^\mathrm{s1}$                     & 1.146 & 13.18 & 0.429 & 23.7 & 
                                          0.898 & 13.75 & 0.354 & 20.5 \\
  CNS                                   & 1.134 & 11.53 & 0.481 & 0 & 
                                          0.890 & 11.58 & 0.410 & 0 \\
   & & & & & & &   \\
  \hline
  \end{tabular}
\end{table*}
The minimum baryonic mass of our non-rotating EPNS sequences is in the
range $0.95 - 1.24 M_{\sun}$. The minimum mass is increased in the
case of Kepler rotating stars to $0.97 - 1.29 M_{\sun})$. If accretion
is neglected the baryonic mass is conserved during the evolution of
the EPNS to the CNS. We follow the evolution of the minimum mass
EPNS$^\mathrm{s5s1}_\mathrm{YL04}$ and
EPNS$^\mathrm{s4s1}_\mathrm{YL04}$ model to a CNS via the LPNS and the
HNS stage (see Table \ref{evolvmin}). The minimum mass of a CNS born
in a supernova is therefore determined by the minimum mass model of
the EPNS sequence. The above ranges of baryonic mass correspond to a
lower limit of the gravitational CNS's mass in the range $0.89 - 1.13
M_{\sun}$. This is by a factor of ten larger than the minimum mass of
the CNS sequence.
This property of EPNS's was also recently found by \cite{GHZ98}.

Variations of the location of the transition region between the envelope 
(high entropy per baryon) and the core (low entropy per baryon) of the 
EPNS's do not change the minimum mass considerably. 
Starting the transition region at lower densities will lead to smaller
minimum masses of the EPNS's, but the lower the initial mass of the 
EPNS is, the higher is the entropy per baryon in the envelope and in 
the core \cite[see e.g.][]{KJM96, PRPLM98}. This effect drives the 
minimum mass back to higher values, so that our results of the 
minimum mass range is a good approximation \cite[see also][]{GHZ98}.

\cite{TWW96} examine the most likely masses of NS's using the
numerical data of \cite{WW95} who simulated Type-II supernovae with
progenitor stars in the mass range between 11 and 40\,$M_{\sun}$.
They obtain a lower limit of the NS mass, which depends on the mass
and the composition of the progenitor, of $1.15-1.27$~$M_{\sun}$. This
lower limit is comparable to our results.

\subsection{Sensitivity of the results}

In Table \ref{neutrinosphere}, we compare the properties of our PNS
and NS models for a fixed baryonic mass, $M_{\rm B}=1.5\,M_{\sun}$. This
canonical value corresponds to the measured gravitational masses,
$1.35 \pm 0.27\,M_{\sun}$, of neutron stars in binary systems
\cite[see][]{TAMT93,VVZ95}.

As can be inferred by comparing the LPNS models
LPNS$^\mathrm{s2}_\mathrm{YL04(64-62)}$,
LPNS$^\mathrm{s2}_\mathrm{YL04(64-22)}$, and
LPNS$^\mathrm{s2}_\mathrm{YL04(64-63)}$ with the model
LPNS$^\mathrm{s2}_\mathrm{YL04}$ (see Sect.~\ref{sec22}), the
location of the neutrino sphere has nearly no effect on the
gravitational mass and the central density. However, the
circumferential radius and the Kepler frequency vary by up to $10\,\%$
and $15\,\%$, respectively.

The use of an isothermal, instead of an isentropic, EOS in the
envelope of the HNS models HNS$^\mathrm{T03s1}$ and
HNS$^\mathrm{T06s2}$ has only small effects on the properties of the
HNS's (see Table \ref{neutrinosphere}). If thermal effects in the
envelope are however neglected (models HNS$^\mathrm{T0s1}$ and
HNS$^\mathrm{T0s2}$), the circumferential radius is reduced by $\sim
10\,\%$. This yields to an increase of the Kepler frequency by $\sim
20\,\%$. Though the assumption of zero temperature in the envelope
does not change the resulting mass and central densities, the error
made in the circumferential radius and in the Kepler frequency might
be rather large.

\section{Discussion and conclusion} \label{sec5}

The aim of this paper was the investigation of the properties of
rapidly rotating PNS's consisting of nuclear matter (n, p, e$^{-}$,
$\mu^{-}$) under the influence of trapped neutrinos.  We used a
recently developed nuclear EOS for hot, dense matter in the nuclear
Thomas-Fermi approach \cite[]{Str98}.  The nuclear EOS was
extended in this paper to subnuclear densities and to different
compositions of PNS and HNS matter (i.e. trapped neutrinos, constant
entropy per baryon, $\ldots$).  Our results for subnuclear densities
are comparable to the results derived by \cite{LS91} used in the
investigations of \cite{GHZ97, GHZ98} and \cite{Gon97}.
However, considerable deviations occur at densities around and
above nuclear matter density,  
since the EOS of \cite{LS91} shows a smaller temperature dependence. 
In general, this difference to our results has only a small
impact on the properties of PNS's, but impacts on HNS's with canonical 
mass. The impact becomes considerable for more massive stars.

An investigation of this kind should investigate the properties of
PNS's and HNS's for the whole range of possible masses.  So far lepton
concentration profiles and entropy per baryon profiles 
were derived only for one fixed mass in simulations of PNS evolution
\cite[e.g.][]{BL86, KJ95, BHF95}.  In our calculation, the extension
to the whole range of masses was done by assuming that the lepton
concentration and the entropy per baryon does not depend on the mass
of the PNS.

As a result we find that the minimum gravitational mass of a NS is
determined at the earliest stage of a PNS, so that the mass of a NS
formed in a Type-II supernova is larger than 0.89 - 1.13\,$M_{\sun}$,  
which confirms similar results of a recent investigation by 
\cite{GHZ98}.
The exact lower limit of the NS mass depends on the used entropy per
baryon in the EPNS model. The quoted mass range was obtained by using
$s = 4$ and $s = 5$ as lower and upper limit of the entropy
per baryon in the envelope of the EPNS \cite[]{BHF95}.  The minimum
mass of EPNS's is approximately by a factor of ten larger than the
minimum stable mass of CNS's.

The maximum possible baryonic mass of a CNS is always larger than the
maximum possible baryonic mass of the PNS's, this means that once a
PNS was formed, it cannot collapse into a black hole, if further
accretion is neglected \cite[see also][]{Tak95, Bom96}.
One exception is the case of the most massive stars 
rotating near or at its Kepler frequency, which possibly 
collapse to a black hole during their time evolution, 
for a discussion, see \cite{CST94} and \cite{SBGH94}.  This
statement holds for stars with a pure nucleonic/leptonic composition.  If
one includes hyperons and/or quarks, the maximum baryonic mass of a
CNS decreases \cite[e.g.][]{HWWS98, BLC99} and may be smaller than the
maximum baryonic mass of a PNS. Then, a sufficiently massive PNS may 
collapse to a black hole during deloptonization \cite[]{Pra97}.  The
maximum gravitational mass increases slightly with increasing entropy
per baryon in contrast to the maximum baryonic mass \cite[see
also][]{Pra97}. As it was pointed out, \cite{GHZ97} got a different
result due to the use of the EOS derived by \cite{LS91}.

It turned out that the influence of rotation has several impacts on
properties of PNS's and NS's.  Whereas the minimum mass changes only
slightly due to rotation, the effect on the maximum mass is rather
large, particularly for CNS's. The central baryon density is nearly
unaffected by rotation at the early stages of the evolution, whereas
the impact on the the later stages is rather large.  The effect on the
circumferential radius decreases with increasing mass and the impact
on EPNS's and LPNS's is slightly larger than the impact on CNS's.

We investigated the influence of different shapes of the neutrino
sphere on the structure of LPNS's. As expected, we obtained considerable
differences (up to 10\,\%) in the circumferential radii and, due to
this, in the Kepler frequency.  Other properties, as the central
density, are barely changed by the location of the neutrino sphere.
We have also considered different temperatures in the envelope of
LPNS's and HNS's.  If a very high temperature of 0.6 MeV is used in
the envelope instead of a constant entropy per baryon, the gross
properties of the LPNS change by less than 1\,\%.  Larger deviations
were obtained if the envelope is assumed to be cold. The finite
temperature effects should therefore not be neglected in the
envelope. Furthermore, we obtain that the thermal effects are
comparable to the effects due to trapped neutrinos. This is in
contrast to the results of \cite{Pra97}, who found the thermal effects
to be much smaller than the effect of high lepton numbers.

In our model we assumed that accretion onto the emerging NS stopped
after the formation of the EPNS \cite[see][]{BHF95}.  Furthermore, we
kept the angular momentum constant during the deleptonization and the
thermal cooling period. Under these presumptions, we obtain a lower
limit on the periods of young NS's with a typical baryonic mass of 1.5
$M_{\sun}$ between 1.56 and 2.22\,ms, which is in accordance with 
similar values obtained by \cite{GHZ97, GHZ98}. These results 
support strongly the hypothesis that millisecond pulsars were 
accelerated due to accretion. With the
same reasoning, one obtains an upper limit of the stability parameter
$\beta\lesssim 0.082$, which is smaller, for almost all NS's, than the 
critical value for the onset of dynamical and secular instabilities. 
Kepler rotating massive CNS's possess stability parameter values up 
to $\beta \sim 0.13$ and might therefore be secular unstable against 
$m=2$ and $m=3$ non-axisymmetric perturbations.

Though we have assumed uniform rotation, it seems to be very
reasonable that PNS's and young NS's rotate differentially
\cite[e.g.][]{JM89}. As it was found by \cite{Schaab98} and \cite{GHZ98},
differential rotation may considerably effect the structure of PNS's
and NS's, and we will pursue our investigations to differential
rotation in the future. Another issue that will be addressed in a
future work is the effect of additional degrees of freedom
(e.g. hyperons) on the evolution of PNS's.

\begin{acknowledgements}

We want to thank Wolfgang Keil, Georg Raffelt, Thomas Strobel and
Fridolin Weber for many helpful discussions.  Two of us, K.~S. and
Ch.~S., gratefully acknowledge the Bavarian State for financial
support.

\end{acknowledgements}

\appendix
\onecolumn
\section{Nuclear equation of state} \label{appendix}

The momentum and density dependent interaction is given by [the upper
(lower) sign corresponds to nucleons with equal (unequal) isospin]
\cite[]{MS90}:
\begin{equation}
\label{A.1}
v_{12\tau} =  - \frac{2T_{0\tau}}{n_0}
\,g\,\left(r_{12}\right) 
\left(\frac{1}{2}(1 \mp \xi) \alpha - \frac{1}{2}(1 \mp \zeta)
\left(\beta \left(\frac{p_{12}}{p_0}\right)^2 - \gamma \frac{p_0}{|p_{12}|}
+ \sigma \left(\frac{2 \bar n}{n_0}\right)^{\frac{2}{3}}\right)\right)~.
\end{equation}
The quantities $n_0$, $p_0$, and $T_{0\tau} (=p_0^2/2m_\tau)$ [$T_0 =
p_0^2/2 \bar m$ with $\bar m = 0.5(m_\mathrm{n} + m_\mathrm{p})$, see
(\ref{A.4})] denote the baryon number density, Fermi momentum, and the
kinetic single-particle energy of symmetric nuclear matter at
saturation ($\tau$ denotes the isospin), respectively.  The choice
$\xi \not= \zeta$ leads to a better description of asymmetric nuclear
systems, and the behaviour of the optical potentials is improved by
the term $\sigma (2\bar n/n_0)^{2/3}$, where $\overline{n}^{2/3} = 0.5
(n_{1}^{2/3} + n_{2}^{2/3})$.

The parameter set is given by the most recent adjustment of
\cite{MS96, MS98}: $\alpha = 1.94684$, $\beta = 0.15311$, $\gamma = 1.13672$,
$\sigma = 1.05$, $\xi = 0.27976$, and $\zeta = 0.55665$, which leads
to the following properties of symmetric nuclear matter at saturation
density ($n_0 = 0.16114$~fm$^{-3}$): energy per baryon $u_0 = -16.24$
MeV, incompressibility $K = 234$ MeV, symmetry energy $J = 32.65$ MeV
and effective nucleon mass $m^*_\tau/m_\tau = 0.867$.

\begin{figure}
  \resizebox{\hsize}{!}{\rotatebox{-90}{\includegraphics{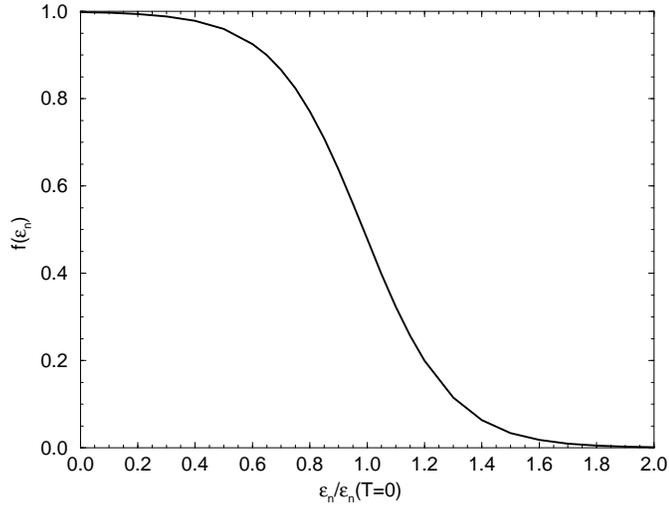}}}
  \caption[]{Fermi-Dirac distribution function for the neutrons of the 
             LPNS$^\mathrm{s2}_\mathrm{YL04}$ EOS at $n = 1$ fm$^{-3}$. 
             $T = 79.8$ MeV, $T_\mathrm{F} = 278.2$ MeV, 
             $T/T_\mathrm{F} \approx 0.29$, $\mu'_\mathrm{n} = 512.8$ MeV 
             and $\epsilon_\mathrm{n}(T=0) = 521.2$ MeV at the Fermi 
             surface.}
  \label{fn}
\end{figure}

The potentials radial dependence, $g$, is chosen to be of Yukawa
type. The function is normalized to unity:
\begin{equation}
\label{A.2}
\int_{-\infty}^{+\infty} g \left(r_{12}\right) \mathrm{d}^{3}r_{1} = 1.
\end{equation}
This leads to the following single particle energy $u$ 
\cite[]{Str98}:
\begin{equation}
\label{A.3}
u = \frac{2}{n (2\pi)^3}\sum_{\tau} \int_{-\infty}^{+\infty} 
       \Bigl(\frac{p_1^2}{2m_{\tau}} 
       + \frac{1}{2} V_{\tau}(p_1) \Bigr)f_{\tau}(p_{1}) \mathrm{d}^{3}p_{1} ,
\end{equation}
with the one-particle potential $V_{\tau}(p_1)$:
\begin{align}
\label{A.4}
V_{\tau}(p_1)  = &-\frac{2}{(2\pi)^3} \frac{2T_{0 \tau}}{n_0}
                   \int_{-\infty}^{+\infty} 
                   \left( \alpha_\mathrm{l} 
                   - \beta_\mathrm{l}\left( \frac{p_{12}}{p_{0}} \right) ^2 
                   + \gamma_\mathrm{l} \frac{p_{0}}{|p_{12}|} 
                   - \sigma_\mathrm{l} \left( \frac{2 \overline{n}}{n_{0}} 
                   \right) ^{\frac{2}{3}} \right) f_{\tau}(p_{2}) 
                   \mathrm{d}^{3}p_{2}
                   \nonumber\\
                     & - \frac{2}{(2\pi)^3} \frac{2T_{0}}{n_0}
                   \int_{-\infty}^{+\infty} 
                   \left( \alpha_\mathrm{u} 
                   - \beta_\mathrm{u}\left( \frac{p_{12}}{p_{0}} \right) ^2 
                   + \gamma_\mathrm{u} \frac{p_{0}}{|p_{12}|} 
                   - \sigma_\mathrm{u} \left( \frac{2 \overline{n}}{n_{0}} 
                   \right) ^{\frac{2}{3}} \right) f_{-\tau}(p_{2}) 
                   \mathrm{d}^{3}p_{2},
\end{align}
[$\tau$ and $-\tau$ denote different isospin and the following
abbreviations were used: $\alpha_\mathrm{l} = 0.5 (1 - \xi) \alpha$,
$\beta_\mathrm{l} = 0.5 (1 - \zeta) \beta$, $\gamma_\mathrm{l} = 0.5
(1 - \zeta) \gamma$, $\sigma_\mathrm{l} = 0.5 (1 - \zeta) \sigma$,
$\alpha_\mathrm{u} = 0.5 (1 + \xi) \alpha$, $\beta_\mathrm{u} = 0.5 (1
+ \zeta) \beta$, $\gamma_\mathrm{u} = 0.5 (1 + \zeta) \gamma$ and
$\sigma_\mathrm{u} = 0.5 (1 + \zeta) \sigma$] and the baryon number
density:
\begin{equation}
\label{A.5}
n = \frac{2}{(2\pi) ^3}\sum_{\tau} \int_{-\infty}^{+\infty} 
    f_{\tau}(p_1) \mathrm{d}^{3}p_{1} .
\end{equation}
In the Eqs. (\ref{A.3}), (\ref{A.4}), and (\ref{A.5})
$f_{\tau}(p)$ denotes the Fermi-Dirac distribution function (see 
Fig.~\ref{fn} as an example) of a
baryon with isospin $\tau$ ($k_\mathrm{B} = 1$):
\begin{equation}
\label{A.6}
   f_{\tau}(p_{1}) = \Bigl( 1 + \exp (\frac{1}{T}(\epsilon_{\tau}(p_{1}) 
                                     -\mu'_{\tau})) \Bigr)^{-1} ,
\end{equation}
where $\epsilon_{\tau}$ denotes the one-particle energy:
\begin{equation}
\label{A.7}
\epsilon_{\tau}(p_{1}) = \frac{p_{1}^{2}}{2m_{\tau}} 
                            + V_{\tau}(p_{1}) .
\end{equation}

{\bf Hint:} $\mu'_{\tau}$ in Eq. (\ref{A.6}) is not the chemical
potential in normal sense, because of the density dependent part in
the interaction, for an explanation see \cite{MS90}, Appendix A. The
chemical potentials of neutrons and protons, $\mu_{\tau}$, can be
derived over the thermodynamic derivatives:
\begin{equation}
\label{A.8}
\mu_{\tau} = \left( n \left( \frac{\partial}{\partial n_\tau}
             \right)_{s,n_{-\tau}} + 1 \right) u + m_\tau
           = \left( n \left( \frac{\partial}{\partial n_\tau}
             \right)_{T,n_{-\tau}} + 1 \right) f + m_\tau ,
\end{equation}
in which $f$ is the free energy per baryon and $m_\tau$ the
rest-mass of neutrons or protons.

\twocolumn

\vspace{\fill}

This article was processed by the author using Springer-Verlag
\LaTeX~A\&A style file L-AA version 4.01.

\end{document}